# A Modified Iterative IOM Approach for Optimization of Biochemical Systems

**Gongxian Xu**[a,*], **Cheng Shao**[a], **Zhilong Xiu**[b]

[a]*Research Center of Information and Control, Dalian University of Technology, Dalian 116024, China*

[b]*Department of Bioscience and Biotechnology, Dalian University of Technology, Dalian 116024, China*

---

**Abstract:** The presented previously indirect optimization method (IOM) developed within biochemical systems theory (BST) provides a versatile and mathematically tractable optimization strategy for biochemical systems. However, due to the local approximations nature of the BST formalism, the iterative version of this technique possibly does not yield the true optimum solution. In this work, an algorithm is proposed to obtain the correct and consistent optimum steady-state operating point of biochemical systems. The existing linear optimization problem of the direct IOM approach is modified by adding an equality constraint of describing the consistency of solutions between the S-system and the original model. Lagrangian analysis is employed to derive the first order necessary optimality conditions for the above modified optimization problem. This leads to a procedure that may be regarded as a modified iterative IOM approach in which the optimization objective function includes an extra linear term. The extra term contains a comparison of metabolite concentration derivatives with respect to the enzyme activities between the S-system and the original model and ensures that the new algorithm is still carried out within linear programming techniques. The presented framework is applied to several biochemical systems and shown to the tractability and effectiveness of the method. The simulation is also studied to investigate the convergence properties of the algorithm and to give a performance comparison of standard and modified iterative IOM approach.

**Keywords:** Optimization; Linear programming; S-system; Lagrangian multiplier; Biochemical systems

---

## 1. Introduction

In recent years, the model-based optimization of biochemical and biotechnological systems has become a crucial component of metabolic engineering. From a technological point of view, mathematical optimization provides a systematic and efficient tool that helps to analyze and optimize these processes to predict the maximum yield or production rate of some desired product. Moreover, by such optimization, it is convenient to obtain some important data about general properties of biochemical systems (Vera et al., 2003a). Once this valuable information is achievable, it will be possible to derive their optimal operation policies of studied biotechnological systems.

Much research has been directed toward the development of model-based optimization strategies, including the mathematical foundations of such approaches (Voit, 1992; Regan et al., 1993; Hatzimanikatis et al., 1996a, 1996b; Torres et al., 1996, 1997; Petkov and Maranas, 1997; Voit and Del Signore, 2001; Torres and Voit, 2002; Marín-Sanguino and Torres, 2003; Vera et al.,

---

*Corresponding author. Tel.: +86-411-84707577; fax: +86-411-84707579.

*E-mail addresses:* dutxugx@yahoo.com.cn (Gongxian Xu), cshao@dlut.edu.cn (Cheng Shao).

2003a; Chang and Sahinidis, 2005) and their application to some processes (Heinrich et al., 1991; Hatzimanikatis et al., 1998; Marín-Sanguino and Torres, 2000, 2002; Alvarez-Vasquez et al., 2000; Vera et al., 2003b; Sevilla et al., 2005). One successful approach to the optimization of biochemical systems is the indirect optimization method (IOM) (Torres et al., 1996, 1997; Voit, 1992; Marín-Sanguino and Torres, 2003; Vera et al., 2003a), which is based on the approximation of the original nonlinear differential equation models describing the biochemical process as an S-system or a GMA system. The S-system models are founded on the Biochemical Systems Theory (BST) introduced by Savageau and co-workers (Savageau 1969a, 1969b, 1970, 1976; Savageau et al., 1987a, 1987b). In this mathematical formalism, the change in each metabolite is represented by two competing power law functions describing aggregation and consumption. The advantage of this representation is that the steady-state equations are linear when the variables of the models are expressed in logarithmic coordinates. This enables the use of linear programming techniques.

When the above-mentioned IOM approach is used to optimize a biochemical system, a possible outcome is that some of the metabolite concentrations exceed significantly the imposed limits or that the original model is unstable. If such a situation occurs, we can apply the iterative IOM version (Marín-Sanguino and Torres, 2000) to find a consistent steady-state. In fact it is a repetition of the direct IOM approach. However, the iterative IOM strategy is strictly valid only near the reference steady-state. The reason for this is that the BST formalism is based on first order Taylor's approximations, which is a local representation of the original system. An example of such a case is the application of the above optimization method to the tryptophan biosynthesis in *Escherichia coli.* (Marín-Sanguino and Torres, 2000). The authors attained a solution similar to the S-system, where the tryptophan flux is more than 3 times the basal flux. But this result is inferior to that of calculating with a rate of tryptophan production increased more than 4 times by the direct IOM approach. Clearly, the former is not a true optimum solution but a local one. To overcome this difficulty of running into a range of local solution and enhance the effectiveness of the iterative IOM approach, it is necessary to make an improvement in its scheme.

For this purpose, in the present study we propose to transform the existing linear optimization problem of the iterative IOM approach into a problem with an additional equality constraint to account for the consistency of solutions between the S-system and the original model. Using the general Lagrangian multiplier method, the resulting optimization problem is modified as an equivalent problem that can be solved with available linear optimization techniques. To demonstrate the validity of the new algorithm, we apply this modified method to three metabolic pathways.

This paper is organized as follows. Section 2 presents the formalism of the standard iterative IOM approach. In section 3, a modified iterative IOM version is developed. Numerical simulations are shown in Section 4, 5 and 6. Finally, brief conclusions are followed in Section 7.

## 2. Standard iterative IOM approach

*2.1 Optimization problem statement*

Consider the following problem of optimizing a biochemical system:

$$\max \quad J(X,Y) \tag{1}$$

subject to satisfying:

$$F_i(X,Y)=0 \qquad i=1,2,\ldots,n \tag{2}$$

$$X_i^L \le X_i \le X_i^U \qquad (3)$$

$$Y_k^L \le Y_k \le Y_k^U \qquad k = 1,2,\ldots,m \qquad (4)$$

$$G_l(X,Y) \le 0 \qquad l = 1,2,\ldots,p \qquad (5)$$

where $X = (X_1, X_2, \ldots, X_n)^T \in R^n$, $Y = (Y_1, Y_2, \ldots, Y_m)^T \in R^m$; the objective function $J$ is usually a flux or a particular metabolite concentration; constraint (2) is the steady-state condition (i.e., $dX_i/dt = 0$); constraint (3) and (4) keep the metabolite concentrations $X_i$ and the enzyme activities $Y_k$ to stay within certain limits; and (5) forces a flux or the ratio of some two fluxes to remain below a certain limit. Due to the product nature of the fluxes, constraint (5) becomes linear in logarithmic space.

*2.2 The direct IOM approach*

The implementation of the method include mainly four steps:

*(1) Translation of the original model to the S-system formalism*

The S-system formalism is based on BST which proposes the use of power law functions to describe the nonlinear nature of biochemical processes (Savageau, 1976). Under this representation, the elementary fluxes consisting of input fluxes and output ones are grouped into aggregate fluxes that pass into and out of metabolic pools. These aggregate fluxes have forms given by "accumulation" flux $V_i^+$ and "consumption" $V_i^-$. Then the original model:

$$\frac{dX_i}{dt} = F_i(X,Y) \qquad i = 1,2,\ldots,n \qquad (6)$$

can be expressed as:

$$\frac{dX_i}{dt} = V_i^+ - V_i^- \qquad i = 1,2,\ldots,n \qquad (7)$$

If each of these rate laws is represented in the power law formalism, then yields the S-system model of Eq. (6):

$$\frac{dX_i}{dt} = \alpha_i \prod_{j=1}^{n} X_j^{g_{ij}} \prod_{k=1}^{m} Y_k^{g'_{ik}} - \beta_i \prod_{j=1}^{n} X_j^{h_{ij}} \prod_{k=1}^{m} Y_k^{h'_{ik}} \qquad i = 1,2,\ldots,n \qquad (8)$$

where the model parameters $g_{ij}$, $g'_{ik}$, $h_{ij}$ and $h'_{ik}$ are the kinetic orders, and $\alpha_i$ and $\beta_i$ are the rate constants. The kinetic orders are defined as:

$$g_{ij} = \left(\frac{\partial V_i^+}{\partial X_j}\frac{X_j}{V_i^+}\right)_0 \qquad g'_{ik} = \left(\frac{\partial V_i^+}{\partial Y_k}\frac{Y_k}{V_i^+}\right)_0 \qquad h_{ij} = \left(\frac{\partial V_i^-}{\partial X_j}\frac{X_j}{V_i^-}\right)_0 \qquad h'_{ik} = \left(\frac{\partial V_i^-}{\partial Y_k}\frac{Y_k}{V_i^-}\right)_0$$

And the rate constants are defined as:

$$\alpha_i = \left(V_i^+\right)_0 \prod_{j=1}^{n} \left(X_j\right)_0^{-g_{ij}} \prod_{k=1}^{m} \left(Y_k\right)_0^{-g'_{ik}} \quad \text{and} \quad \beta_i = \left(V_i^-\right)_0 \prod_{j=1}^{n} \left(X_j\right)_0^{-h_{ij}} \prod_{k=1}^{m} \left(Y_k\right)_0^{-h'_{ik}}$$

where the subscript 0 indicates that the results are evaluated at the steady-state of metabolite concentrations. Based on Eq. (8) the objective function $J(X,Y)$ can also be written as the following S-system form:

$$J'(X,Y) = \gamma \prod_{i=1}^{n} X_i^{f_i} \prod_{k=1}^{m} Y_k^{f'_k} \qquad (9)$$

In Eq. (9) $f_i$ and $f'_k$ terms stand for the kinetic orders, and $\gamma$ represents the corresponding rate constant.

*(2) Quality assessment of the S-system model*

The S-system formalism has a significant advantage in that it facilitates the analytical and numerical quality assessment (such as with the software package PLAS, Ferreira, 2000). Firstly, it allows us to detect the local stability of the steady-state, which can be computed by solving the characteristic equation of the following matrix (Savageau, 1976; Chen, 1984):

$$\begin{bmatrix} d_{11} & d_{12} & \cdots & d_{1n} \\ d_{21} & d_{22} & \cdots & d_{2n} \\ \vdots & \vdots & \ddots & \vdots \\ d_{n1} & d_{n2} & \cdots & d_{nn} \end{bmatrix} \tag{10}$$

where

$$d_{ij} = \left(\frac{V_i^+}{X_i}\right)_0 \left(g_{ij} - h_{ij}\right)$$

If all real parts of the eigenvalues are negative, then the steady-state is locally stable.

Secondly, the robustness analysis of the model can be done, indicating whether the model is able to tolerate small structural changes. The system sensitivity theory provides important methods for characterizing the quality of a model. There are three types of sensitivity coefficients, which are defined as follows.

*Rate constant sensitivities*

The rate constant sensitivities are defined as the ratio of the percentage change in a systemic variable to an infinitesimal percentage change in a rate constant:

$$S(X_i, \alpha_j) = \left(\frac{\partial X_i}{\partial \alpha_j}\frac{\alpha_j}{X_i}\right)_0 \quad S(X_i, \beta_j) = \left(\frac{\partial X_i}{\partial \beta_j}\frac{\beta_j}{X_i}\right)_0$$

$$S(V_i, \alpha_j) = \left(\frac{\partial V_i}{\partial \alpha_j}\frac{\alpha_j}{V_i}\right)_0 \quad S(V_i, \beta_j) = \left(\frac{\partial V_i}{\partial \beta_j}\frac{\beta_j}{V_i}\right)_0$$

where $V_i$ represents a given flux. These sensitivities can be calculated by differentiation of the explicit solution (Voit, 2000; Torres and Voit, 2002). It is easy to know that these sensitivities are only dependent upon the kinetic orders of the system. So they are properties of the integrated system and not its isolated components.

*Kinetic order sensitivities*

A kinetic order sensitivity coefficient is defined as the ratio of the percentage change in a systemic variable to an infinitesimal percentage change in a kinetic order, $g_{ij}$, $g'_{ik}$, $h_{ij}$ or $h'_{ik}$:

$$S(X_i, g_{qj}) = \left(\frac{\partial X_i}{\partial g_{qj}}\frac{g_{qj}}{X_i}\right)_0 \quad S(V_i, g_{qj}) = \left(\frac{\partial V_i}{\partial g_{qj}}\frac{g_{qj}}{V_i}\right)_0$$

$$S(X_i, g'_{qk}) = \left(\frac{\partial X_i}{\partial g'_{qk}}\frac{g'_{qk}}{X_i}\right)_0 \quad S(V_i, g'_{qk}) = \left(\frac{\partial V_i}{\partial g'_{qk}}\frac{g'_{qk}}{V_i}\right)_0$$

$$S(X_i, h_{qj}) = \left(\frac{\partial X_i}{\partial h_{qj}}\frac{h_{qj}}{X_i}\right)_0 \quad S(V_i, h_{qj}) = \left(\frac{\partial V_i}{\partial h_{qj}}\frac{h_{qj}}{V_i}\right)_0$$

$$S(X_i, h'_{qk}) = \left(\frac{\partial X_i}{\partial h'_{qk}}\frac{h'_{qk}}{X_i}\right)_0 \quad S(V_i, h'_{qk}) = \left(\frac{\partial V_i}{\partial h'_{qk}}\frac{h'_{qk}}{V_i}\right)_0$$

where $q = 1, 2, \ldots, n$. Again, these sensitivities are properties of the integrated system and not its isolated components, but here the sensitivities are a function of both rate constants and kinetic orders. The kinetic order sensitivities can also be calculated by differentiation of the explicit solution (Voit, 2000; Torres and Voit, 2002). In a good model the sensitivities must be small, otherwise high sensitivities (i.e., absolute values upper than 50) (Vera et al., 2003b) indicate that the model is ill-determined. Once such a bad case happens, the portions of investigated model need to be given a more attention.

*Logarithmic Gains*

The logarithmic gains are specific types of sensitivity coefficients that are defined as:

$$L(X_i, Y_k) = \left( \frac{\partial X_i}{\partial Y_k} \frac{Y_k}{X_i} \right)_0 \quad \text{and} \quad L(V_i, Y_k) = \left( \frac{\partial V_i}{\partial Y_k} \frac{Y_k}{V_i} \right)_0$$

The former is called concentration logarithmic gains and the latter is called flux logarithmic gains. Like the previous sensitivities, the logarithmic gains should have low values (less than 10 in absolute value) (Vera et al., 2003b).

Thirdly, we can check the dynamic features that characterize the transient responses to temporary perturbations or permanent alterations. These considerations include: How long does it take for the system to return to the steady-state after a given increase or decrease in intermediate metabolites and enzyme activities? How will the model respond to a medium-sized but not slight perturbation? Whether the perturbation leads to a response in the form of oscillations? If that is the truth, are they observed in the real system? Whether the transient responses are reasonable? Such analyses often identify problems of consistency and reliability of the mathematical representation (Shiraishi and Savageau, 1992; Ni and Savageau, 1996a; Ni and Savageau, 1996b).

*(3) Linear programming and optimization*

Although S-system models are nonlinear, the steady-state equations are linear when the variables are expressed in logarithmic coordinates (Savageau, 1969b). This allows us to use the linear programming techniques (Voit, 1992; Regan et al., 1993; Torres et al., 1996).

At steady-state the S-system (8) reduces to the following nonlinear equations:

$$\sum_{j=1}^{n} (g_{ij} - h_{ij}) \ln(X_j) + \sum_{k=1}^{m} (g'_{ik} - h'_{ik}) \ln(Y_k) = \ln\left( \frac{\beta_i}{\alpha_i} \right) \qquad i = 1,2,\ldots,n \tag{11}$$

Let $x_j = \ln(X_j), j = 1,2,\ldots,n, y_k = \ln(Y_k), k = 1,2,\ldots,m$ and $b_i = \ln(\beta_i / \alpha_i), i = 1,2,\ldots,n$, then Eq. (11) can be recast in a form of linear algebraic equations:

$$A_d x + A_{id} y = b \tag{12}$$

where the matrixes $A_d = (g_{ij} - h_{ij})_{n\times n}$, $A_{id} = (g'_{ik} - h'_{ik})_{n\times m}$, the vectors $x = (x_1, x_2, \cdots, x_n)^T$, $y = (y_1, y_2, \cdots, y_m)^T$ and $b = (b_1, b_2, \cdots, b_n)^T$. If the matrix $A_d$ is non-singular, then $x$ can be solved by Eq. (12):

$$x(y,b) = -A_d^{-1} A_{id} y + A_d^{-1} b \tag{13}$$

**Remark 1.** The inverse of the matrix $A_d$ exists if the system has a non-zero steady-state point (Savageau, 1976).

Due to the fact that the logarithmic transformation does not change the locations of maximum of a function, the nonlinear optimization problem in section 2.1 can be transformed to the following linear programming formulations:

$$\max \quad \bar{J}(x, y)$$

subject to satisfying:

$$A_d x + A_{id} y = b$$

$$\ln(X_i^L) \le x_i \le \ln(X_i^U) \qquad i = 1,2,\dots,n \qquad (14)$$
$$\ln(Y_k^L) \le y_k \le \ln(Y_k^U) \qquad k = 1,2,\dots,m$$
$$\overline{G}(x,y) \le 0$$

where the vector function $\overline{G}(x,y) \in R^l$ is the linear representations of constraint (5) in logarithmic space, and the new objective function $\overline{J}(x,y)$ can be expressed as:

$$\begin{aligned}\overline{J}(x,y) &= \ln(J(X,Y)) \\ &= \ln(J^{'}(X,Y)) \\ &= \ln(\gamma) + \sum_{i=1}^{n} f_i \ln(X_i) + \sum_{k=1}^{m} f_k^{'} \ln(Y_k) \\ &= \ln(\gamma) + \sum_{i=1}^{n} f_i x_i + \sum_{k=1}^{m} f_k^{'} y_k \\ &= \ln(\gamma) + f^T x + f^{'T} y \qquad (15)\end{aligned}$$

where $f = (f_1, f_2, \dots, f_n)^T$, $f^{'} = (f_1^{'}, f_2^{'}, \dots, f_m^{'})^T$.

Note that, to ensure that the optimum solution is within the physiologically acceptable range of values, the following relations are imposed:

$$X_i^L = 0.8(X_i)_0 \text{ and } X_i^U = 1.2(X_i)_0 \qquad (16)$$

where $(X_i)_0$ is the basal steady-state of $X_i$.

*(4) Transfer of results to the original model*

The S-system (8) has been derived as an approximation of the model (6), and it is interesting to explore to what degree any optimized solution is consistent between the two models. To do this, after substituting the enzyme activities $Y_k$ of the optimized S-solution into Eq. (6), the metabolite concentrations $X_i$ are uniquely specified. Since the result is computed via S-system approximation, it is possibly an approximate optimum of the original optimization problem. Still the differences between the steady-states of the original and the S-system models are often small in comparison to the experimental accuracy and in light of other uncertainties involved in any modeling effort (Torres et al., 1996). A possible outcome of the present step is that some of the metabolite concentrations exceed the imposed limits or that Eq. (6) is unstable. In these cases, some of the constraints in step (4) must be changed accordingly.

*2.3 The iterative IOM approach*

When such a situation occurs or significant discrepancies between the S-system and the original model are detected, an efficient procedure can be applied (Voit, 1992; Marín-Sanguino and Torres, 2000) to obtain a consistent steady-state, which is an iterative process of the IOM approach. Each iteration will find a new steady-state. Eventually, this procedure will finish until a satisfactory result is achieved.

## 3. Modified iterative IOM approach

Since the S-system formalism is a local description of the original system at a basal steady-state based on first order Taylor’s approximations, the iterative IOM strategy is strictly effective around this steady-state. It has been shown (Marín-Sanguino and Torres, 2000) that the above approach will not achieve the correct optimum steady-state. Although the authors attained a solution similar to the S-system, where the tryptophan flux is more than 3 times the basal flux, this result is lower than that of finding with a rate of tryptophan production increased more than 4 times by the direct IOM approach. In this study, we propose a modification scheme of this strategy to improve the effectiveness of the iterative IOM approach. This proposed framework just modifies the linear optimization problem in section 2.2 grounded on the similar thought of integrated system optimization and parameter estimation (ISOPE) (Roberts, 1995).

Assume the system (6) has a non-zero steady-state point, denoted as $\hat{X}_i(Y)(i=1,2,\ldots,n)$. Let $\hat{x}_i=\ln(\hat{X}_i)$ and $\hat{x}=(\hat{x}_1,\hat{x}_2,\ldots,\hat{x}_n)^T$. Define $\tilde{J}(y,b)=\bar{J}(x(y,b),y)$. Now let us introduce an additional variable $w\in R^m$, and add the constraint $\hat{x}(y)=x(y,b)$ into the linear optimization problem (14) in section 2.2. Then we have the following modified form of problem (14):

$\min \quad -\tilde{J}(y,b)$

subject to satisfying:

$$\begin{aligned}
&x(w,b)=\hat{x}(w)\\
&x^l\le x(y,b)\le x^u\\
&y^l\le y\le y^u\\
&\tilde{G}(y,b)\le 0\\
&w=y
\end{aligned}\qquad (17)$$

where $\tilde{G}(y,b)=\bar{G}(x(y,b),y)$

$$x^l=(\ln(X_1^L),\ln(X_2^L),\ldots,\ln(X_n^L))^T$$

$$x^u=(\ln(X_1^U),\ln(X_2^U),\ldots,\ln(X_n^U))^T$$

$$y^l=(\ln(Y_1^L),\ln(Y_2^L),\ldots,\ln(Y_m^L))^T$$

$$y^u=(\ln(Y_1^U),\ln(Y_2^U),\ldots,\ln(Y_m^U))^T$$

The optimization problem (17) can then be considered as that of determining the stationary point of the Lagrangian function:

$$\begin{aligned}
L_a(\lambda,\sigma,\mu_1,\mu_2,\eta_1,\eta_2,\eta_3)=&-\tilde{J}(y,b)+\lambda^T(w-y)+\sigma^T[x(w,b)-\hat{x}(w)]+\mu_1^T(y-y^u)+\mu_2^T(-y+y^l)\\
&+\eta_1^T(x(y,b)-x^u)+\eta_2^T(-x(y,b)+x^l)+\eta_3^T\tilde{G}(y,b)
\end{aligned}\qquad (18)$$

where $\lambda$, $\sigma$, $\mu_1$, $\mu_2$, $\eta_1$, $\eta_2$ and $\eta_3$ are Lagrangian multiplier vectors.

Assuming that the required derivatives exist and are continuous, the necessary optimality conditions for the modified optimization problem (17) are:

$$\frac{\partial^T L_a}{\partial y}=-\frac{\partial^T \tilde{J}(y,b)}{\partial y}-\lambda+\mu_1-\mu_2+\frac{\partial^T x(y,b)}{\partial y}(\eta_1-\eta_2)+\frac{\partial^T \tilde{G}(y,b)}{\partial y}\eta_3=0 \qquad (19)$$

$$\frac{\partial^T L_a}{\partial w}=\lambda+\left[\frac{\partial x(w,b)}{\partial w}-\frac{\partial \hat{x}(w)}{\partial w}\right]^T\sigma=0 \qquad (20)$$

$$\frac{\partial^T L_a}{\partial b}=-\frac{\partial^T \tilde{J}(y,b)}{\partial b}+\frac{\partial^T x(w,b)}{\partial b}\sigma+\frac{\partial^T x(y,b)}{\partial b}(\eta_1-\eta_2)+\frac{\partial^T \tilde{G}(y,b)}{\partial b}\eta_3=0 \qquad (21)$$

$$\frac{\partial^T L_a}{\partial \sigma}=x(w,b)-\hat{x}(w)=0 \qquad (22)$$

$$\frac{\partial^T L_a}{\partial \lambda}=w-y=0 \qquad (23)$$

$$y-y^u\le 0,\ \mu_1\ge 0,\ \mu_1^T(y-y^u)=0 \qquad (24)$$

$$-y+y^l\le 0,\ \mu_2\ge 0,\ \mu_2^T(-y+y^l)=0 \qquad (25)$$

$$x(y,b)-x^u \le 0,\ \eta_1 \ge 0,\ \eta_1^T(x(y,b)-x^u)=0 \tag{26}$$

$$-x(y,b)+x^l \le 0,\ \eta_2 \ge 0,\ \eta_2^T(-x(y,b)+x^l)=0 \tag{27}$$

$$\tilde{G}(y,b)\le 0, \eta_3 \ge 0, \eta_3^T\tilde{G}(y,b)=0 \tag{28}$$

From Eq. (20) and (21), eliminating the Lagrangian multiplier $\sigma$, we obtain the formula for the Lagrangian multiplier $\lambda$:

$$\lambda(w,b)=\left[\frac{\partial x(w,b)}{\partial w}-\frac{\partial \hat{x}(w)}{\partial w}\right]^T\left[\frac{\partial^T x(w,b)}{\partial b}\right]^{-1}\left[-\frac{\partial^T \tilde{J}(y,b)}{\partial b}+\frac{\partial^T x(y,b)}{\partial b}(\eta_1-\eta_2)+\frac{\partial^T \tilde{G}(y,b)}{\partial b}\eta_3\right] \tag{29}$$

and the necessary optimality conditions for the problem (17) can be reduced to a set of equations including (19) and (22)-(28).

By Eq. (13) and (23), we have

$$\frac{\partial x(y,b)}{\partial b}=A_d^{-1} \tag{30}$$

$$\frac{\partial x(w,b)}{\partial w}=-A_d^{-1}A_{id} \tag{31}$$

$$\frac{\partial x(w,b)}{\partial b}=A_d^{-1} \tag{32}$$

Using now the relations

$$\frac{\partial \tilde{J}(y,b)}{\partial b}=\frac{\partial \bar{J}(x(y,b),y)}{\partial x}\frac{\partial x(y,b)}{\partial b}=f^T A_d^{-1} \tag{33}$$

$$\frac{\partial \tilde{G}(y,b)}{\partial b}=\frac{\partial \bar{G}(x(y,b),y)}{\partial x}\frac{\partial x(y,b)}{\partial b}=\frac{\partial \bar{G}(x(y,b),y)}{\partial x}A_d^{-1} \tag{34}$$

and putting (30), (31) and (32) into (29) we get the further formula for the Lagrangian multiplier $\lambda$:

$$\lambda(w,b)=\left[-A_d^{-1}A_{id}-\frac{\partial \hat{x}(w)}{\partial w}\right]^T A_d^T\left[-\left(A_d^{-1}\right)^T f+\left(A_d^{-1}\right)^T(\eta_1-\eta_2)+\left(A_d^{-1}\right)^T\frac{\partial^T \bar{G}(x(y,b),y)}{\partial x}\right]$$

$$=\left[A_d^{-1}A_{id}+\frac{\partial \hat{x}(w)}{\partial w}\right]^T\left[f-\eta_1+\eta_2-\frac{\partial^T \bar{G}(x(y,b),y)}{\partial x}\right] \tag{35}$$

Let us now define the following modified optimization problem:

$$\min_y \quad \left\{-\tilde{J}(y,b)-\lambda^T(w,b)y\right\}$$

subject to satisfying:

$$x^l \le x(y,b) \le x^u \tag{36}$$

$$y^l \le y \le y^u$$

$$\tilde{G}(y,b)\le 0$$

Notice that (19) and (24)-(28) are precisely the necessary optimality conditions for the modified optimization problem. Comparing with problem (14), an extra linear term that contains a comparison of metabolite concentration derivatives with respect to the enzyme activities between the S-system and the original model is introduced in (36). This variant will also keep the modified optimization problem solved with available linear programming techniques.

Now we summarize the modified iterative IOM algorithm presented in this paper.

*Step 0*. Given a stable and robust steady-state point $(X^0,Y^0)$, initial multiplier $\eta_1^0$, $\eta_2^0$ and $\eta_3^0$, $\eta_1^0,\eta_2^0,\eta_3^0 \ge 0$, relaxation coefficients $\theta_1$, $\theta_2$, $\theta_3$ and $\theta_4$, $0<\theta_1\le 1$, $\theta_2,\theta_3,\theta_4>0$, some solution accuracy $\varepsilon$, $\varepsilon>0$. Set $r=0$.

*Step 1*. Apply $Y^r$ to the system (6) and find the concentrations $X^r$. Transfer system (6) and objective function $J(X,Y)$ to the S-system forms.

*Step 2*. Perform quality assessments of the S-system model. If it is a valid model, then go to *Step 3*. Else return to *Step 1* and modify $Y^r$.

*Step 3*. For $w = w^r$, $b = b^r$ and $\lambda(w,b) = \lambda(w^r, b^r)$, solve the modified optimization problem (36). Let $\hat{y}^r = \hat{y}^r(w^r, b^r, \eta_1^r, \eta_2^r, \eta_3^r)$ be the solution, with the corresponding multipliers $\hat{\mu}_1^r$, $\hat{\mu}_2^r$, $\hat{\eta}_1^r$, $\hat{\eta}_2^r$ and $\hat{\eta}_3^r$. Denote $\hat{Y}^r = \left(\exp(\hat{y}_1^r), \exp(\hat{y}_2^r), \cdots, \exp(\hat{y}_m^r)\right)^T$.

*Step 4*. If $\left\|\hat{Y}^r - Y^r\right\| \le \varepsilon$, $\left\|\hat{\eta}_1^r - \eta_1^r\right\| \le \varepsilon$, $\left\|\hat{\eta}_2^r - \eta_2^r\right\| \le \varepsilon$ and $\left\|\hat{\eta}_3^r - \eta_3^r\right\| \le \varepsilon$, then transfer the results to the original model and stop.

*Step 5*. Update enzyme activities and multiplies:

$$Y^{r+1} = Y^r + \theta_1(\hat{Y}^r - Y^r) \tag{37}$$

$$\eta_{1i}^{r+1} = \max[0, \eta_{1i}^r + \theta_2(\hat{\eta}_{1i}^r - \eta_{1i}^r)], \qquad i = 1,2,\ldots,n \tag{38}$$

$$\eta_{2i}^{r+1} = \max[0, \eta_{2i}^r + \theta_3(\hat{\eta}_{2i}^r - \eta_{2i}^r)], \qquad i = 1,2,\ldots,n \tag{39}$$

$$\eta_{3l}^{r+1} = \max[0, \eta_{3l}^r + \theta_4(\hat{\eta}_{3l}^r - \eta_{3l}^r)], \qquad l = 1,2,\ldots,p \tag{40}$$

Set $r = r+1$ and continue from *Step 1*.

**Remark 2.** In the practical implementation of the algorithm *Step 2* can be considered until the condition $\left\|\hat{Y}^r - Y^r\right\| \le \varepsilon$ is held.

**Remark 3.** Notice that all elements of the multiplier vectors $\eta_1^r$, $\eta_2^r$ and $\eta_3^r$ are updated with (38)-(40) respectively in order to assure that $\eta_1, \eta_2, \eta_3 \ge 0$. The value of parameter $\theta_1$ is generally selected as 1.

**Remark 4.** When using the modified iterative IOM approach to optimize a system with multiple steady-states, it is necessary to check whether multiple steady-states are obtained after a round of IOM. If this situation occurs, one should choose the stable and robust steady-state that maximizes the performance index as the basal candidate of next IOM.

## 4. Case study 1: Tryptophan biosynthesis in Escherichia coli

*4.1 Optimization of tryptophan biosynthesis in Escherichia coli.*

To verify the calculation algorithm, we will first apply the proposed method to tryptophan biosynthesis in *Escherichia coli*. A schematic network of the simplified metabolic pathway is depicted graphically in Fig. 1. A complete description of the metabolic pathway can be found in Xiu et al. (1997). In this work the mathematical model considers both feedback inhibition of the biosynthetic enzymes and repression of the *trp* operon by tryptophan and explicitly takes into account the growth rate and the demand of tryptophan for protein synthesis. The differential equations in dimensionless variables are given as:

$$\frac{dX_1}{dt} = \frac{X_3 + 1}{1 + (1 + Y_2)X_3} - (Y_8 + Y_1)X_1 \tag{41}$$

$$\frac{dX_2}{dt} = X_1 - (Y_9 + Y_1)X_2 \tag{42}$$

$$\frac{dX_3}{dt} = \frac{X_2 Y_3^2}{Y_3^2 + X_3^2} - (Y_{10} + Y_1)X_3 - \frac{X_3 Y_4}{1 + X_3} - \frac{Y_5(1 - Y_6 Y_1)Y_1 X_3}{X_3 + Y_7} \tag{43}$$

Here, $X_1$ is used for mRNA concentration, $X_2$ is used for enzyme concentration and $X_3$ is used for tryptophan concentration. As we need positive variables to make the required logarithmic transformation, $Y_6$ will be taken as positive and it will be preceded by a minus sign in Xiu's model.

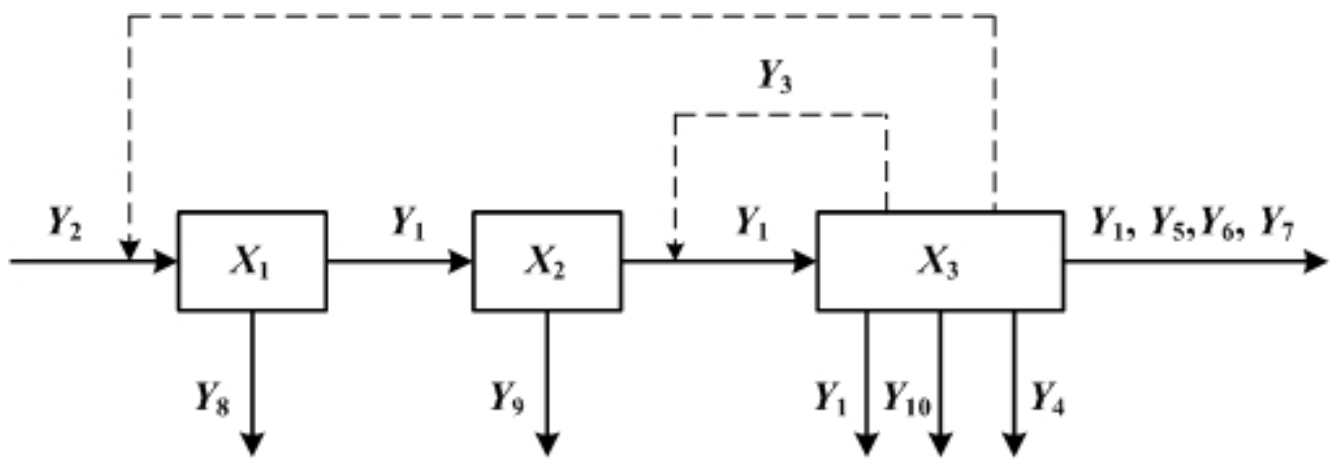

Fig. 1. Diagram of the simplified pathway for tryptophan biosynthesis.

The above-mentioned model does not explicitly account for the tryptophan rate production, but the last term of the right hand of Eq. (43), which is an accumulative term accounting for both consumption and secretion of tryptophan, can be selected as the objective function. This leads to the following optimization problem (Marín-Sanguino and Torres, 2000):

$$\max \quad J = \frac{Y_5(1-Y_6Y_1)Y_1X_3}{X_3+Y_7}$$

subject to satisfying:

$$\begin{aligned}
&\frac{X_3+1}{1+(1+Y_2)X_3} = (Y_8+Y_1)X_1 \\
&X_1 = (Y_9+Y_1)X_2 \\
&\frac{X_2Y_3^2}{Y_3^2+X_3^2} = (Y_{10}+Y_1)X_3 + \frac{X_3Y_4}{1+X_3} + \frac{Y_5(1-Y_6Y_1)Y_1X_3}{X_3+Y_7} \\
&0.8X_{i0} \le X_i \le 1.2X_{i0} \qquad i=1,2,3 \\
&0 < Y_1 \le 0.00624 \\
&4 \le Y_2 \le 10 \\
&500 \le Y_3 \le 5000 \\
&Y_4 = 0.0022Y_2 \\
&0 < Y_5 \le 1000 \\
&(Y_6, Y_7, Y_8, Y_9, Y_{10}) = (7.5, 0.005, 0.9, 0.02, 0)
\end{aligned} \tag{44}$$

By Xiu et al. (1997), there is a unique positive steady-state solution satisfying Eq. (41), (42) and (43), which can be expressed as:

$$X_1 = \frac{1}{Y_8+Y_1}\frac{1+X_3}{1+(1+Y_2)X_3} \tag{45}$$

$$X_2 = \frac{1}{Y_8+Y_1}\frac{1}{Y_9+Y_1}\frac{1+X_3}{1+(1+Y_2)X_3} \tag{46}$$

$$\frac{1}{Y_8+Y_1}\frac{1}{Y_9+Y_1}\frac{1+X_3}{1+(1+Y_2)X_3}\frac{Y_3^2}{Y_3^2+X_3^2} = (Y_{10}+Y_1)X_3 + \frac{X_3Y_4}{1+X_3} + \frac{Y_5(1-Y_6Y_1)Y_1X_3}{X_3+Y_7} \tag{47}$$

Given a set of fixed parameters, $X_3$ is uniquely determined by Eq. (47).

At the basal steady-state (see Table 1), the dynamical model of the pathway is first transformed into an S-system and studied for optimization of tryptophan production by Marín-Sanguino and Torres (2000). Here the S-system representation is modified slightly as:

$$\frac{dX_1}{dt} = 0.6403X_3^{-5.87\times10^{-4}}Y_2^{-0.8332} - 1.0233X_1Y_1^{0.0035}Y_8^{0.9965} \tag{48}$$

$$\frac{dX_2}{dt} = X_1 - 1.4854X_2Y_1^{0.1349}Y_9^{0.8651} \tag{49}$$

$$\frac{dX_3}{dt} = 0.5534X_2X_3^{-0.5573}Y_3^{0.5573} - 1.7094X_3^{0.7684}Y_1^{0.9904}Y_4^{0.0042}Y_5^{0.2274}Y_6^{-5.45\times110^{-3}}Y_7^{-0.8\times10^{-6}} \tag{50}$$

where the parameter $Y_{10}$ is omitted.

Since the objective function $J$ does not include the variables $X_1$ and $X_2$, both the first and the second column of $f$ equal to zero. This implies that the Lagrangian multiplier $\lambda$ has the following formalism:

$$\lambda(w,b)=\left[A_d^{-1}A_{id}+\frac{\partial\hat{x}(w)}{\partial w}\right]^T\left[(0,0,f_3)^T-\eta_1+\eta_2\right] \tag{51}$$

where $\partial\hat{x}(w)/\partial w$ can be obtained from $w=y$ and the following relation:

$$\frac{\partial\hat{x}_i(y_k)}{\partial y_k}=\frac{\partial\hat{X}_i}{\partial Y_k}\frac{Y_k}{\hat{X}_i}$$

where $i=1,2,3$, $k=1,2,\ldots 5$.

*4.2 Performance of the standard iterative IOM approach*

Figs. 2 and 3 show the corresponding variation in enzyme activities $Y_1$ and $Y_2$, metabolite concentrations $X_1$, $X_2$ and $X_3$, and optimized flux $J$ during the standard iterative IOM approach. From Fig. 3, we can see that the standard iterative IOM strategy yields an optimum steady-state solution with an objective index increased less than four times its basal value (see Table 1). Although the obtained steady-state is robust enough and stable (results not shown), the final optimization flux is smaller than the one attained by using the direct IOM approach (see Table 2). These results clearly show the unsatisfactory behavior of the standard iterative IOM approach when applied to the present problem. The method finds an approximately consistent steady-state except for the variable $X_3$, but fails to determine the correct optimum solution.

*4.3 Performance of the modified iterative IOM approach*

Figs. 4 and 5 illustrate the corresponding variation of enzyme activities, metabolite concentrations and optimization index during the modified iterative IOM approach. The relaxation coefficients $\theta_1$, $\theta_2$ and $\theta_3$ are selected as 0.9, 0.8 and 0.8 respectively. It can be observed that the method shows a rapid convergence behavior and produces a higher rate of tryptophan production than the standard iterative IOM approach. The optimized results within 7 iterations are given in Table 3. Compared with the direct and standard iterative IOM approach, the only differences in parameter values are detected in growth rate $Y_1$ and inhibition constant $Y_3$. However, unlike the former two approaches, the modified iterative IOM algorithm not only eventually converges to the correct optimum steady-state, but also achieves the consistent S-system and IOM solutions.

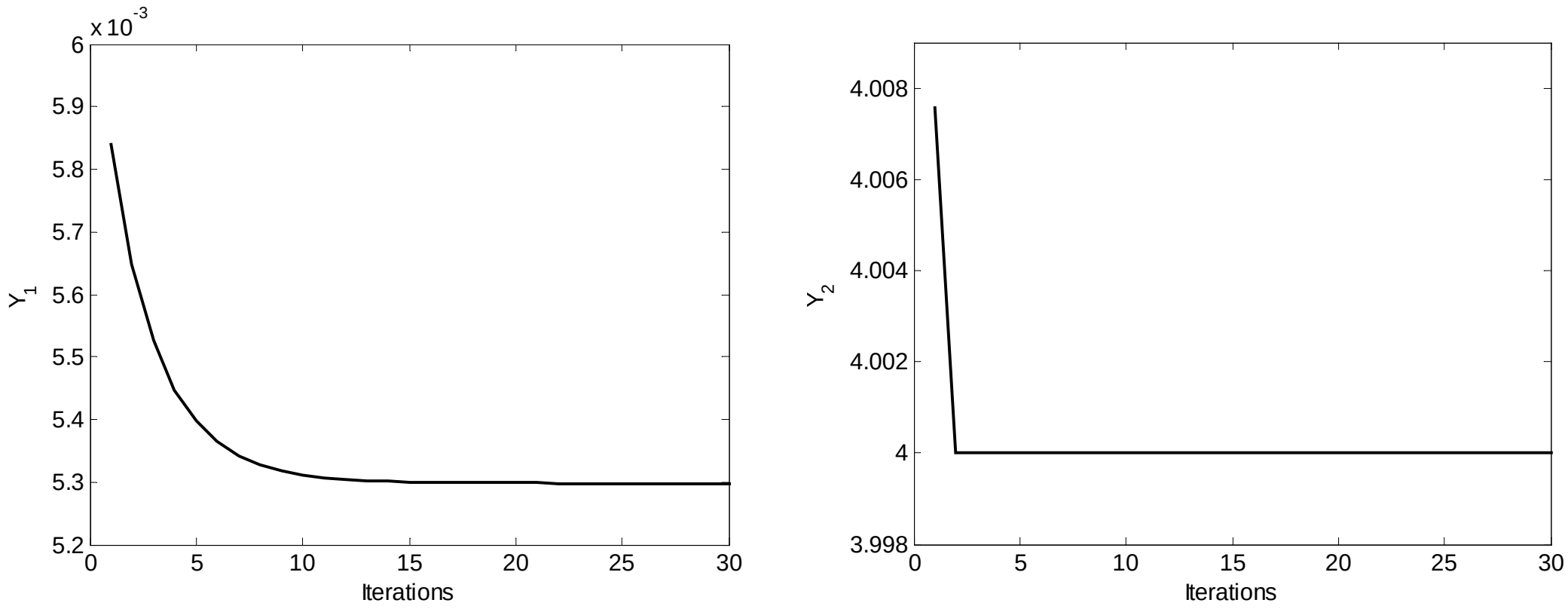

Fig. 2. Variation of enzyme activities for Case study 1 during the standard iterative IOM approach.

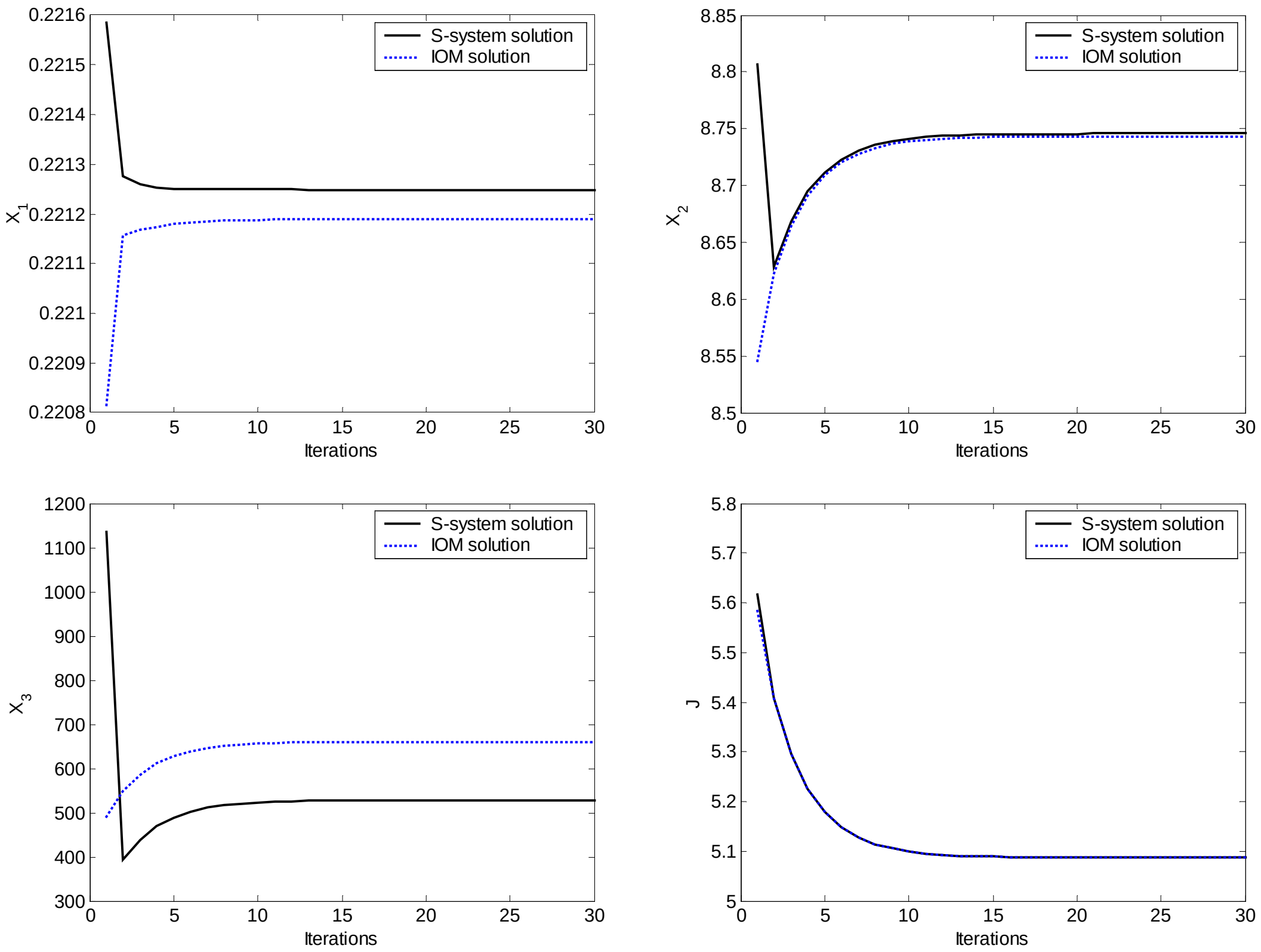

Fig. 3. Variation of metabolite concentrations and optimization index for Case study 1 during the standard iterative IOM approach.

Table 1

Optimal solutions of Case study 1 obtained by using the standard iterative IOM approach

| Variables | Basal steady-state | Optimized solutions (20 iterations) | |
|---|---|---|---|
| | | S-system | IOM |
| $X_1$ | 0.184654 | 1.198 $(X_1)_0$ | 1.198 $(X_1)_0$ |
| $X_2$ | 7.986756 | 1.095 $(X_2)_0$ | 1.095 $(X_2)_0$ |
| $X_3$ | 1418.931944 | 0.372 $(X_3)_0$ | 0.465 $(X_3)_0$ |
| $Y_1$ | 0.00312 | 0.0053 | 0.0053 |
| $Y_2$ | 5 | 4 | 4 |
| $Y_3$ | 2283 | 5000 | 5000 |
| $Y_5$ | 430 | 1000 | 1000 |
| $J$ | 1.310202 | 3.883 $(J)_0$ | 3.883 $(J)_0$ |

Table 2

Optimal solutions of Case study 1 obtained by using the direct IOM approach

| Variables | Basal steady-state | Optimized solutions | |
|---|---|---|---|
| | | S-system | IOM |
| $X_1$ | 0.184654 | 1.200 $(X_1)_0$ | 1.196 $(X_1)_0$ |
| $X_2$ | 7.986756 | 1.103 $(X_2)_0$ | 1.070 $(X_2)_0$ |
| $X_3$ | 1418.931944 | 0.800 $(X_3)_0$ | 0.347 $(X_3)_0$ |
| $Y_1$ | 0.00312 | 0.00584 | 0.00584 |
| $Y_2$ | 5 | 4.008 | 4.008 |
| $Y_3$ | 2283 | 5000 | 5000 |
| $Y_5$ | 430 | 1000 | 1000 |
| $J$ | 1.310202 | 4.287 $(J)_0$ | 4.261 $(J)_0$ |

Table 3
Optimal solutions of Case study 1 obtained by using the modified iterative IOM approach

| Variables | Basal steady-state | Optimized solutions (7 iterations) | |
|---|---|---|---|
| | | S-system | IOM |
| $X_1$ | 0.184654 | $1.198\,(X_1)_0$ | $1.198\,(X_1)_0$ |
| $X_2$ | 7.986756 | $1.055\,(X_2)_0$ | $1.055\,(X_2)_0$ |
| $X_3$ | 1418.931944 | $0.273\,(X_3)_0$ | $0.273\,(X_3)_0$ |
| $Y_1$ | 0.00312 | 0.00624 | 0.00624 |
| $Y_2$ | 5 | 4 | 4 |
| $Y_3$ | 2283 | 4992 | 4992 |
| $Y_5$ | 430 | 1000 | 1000 |
| $J$ | 1.310202 | $4.54\,(J)_0$ | $4.54\,(J)_0$ |

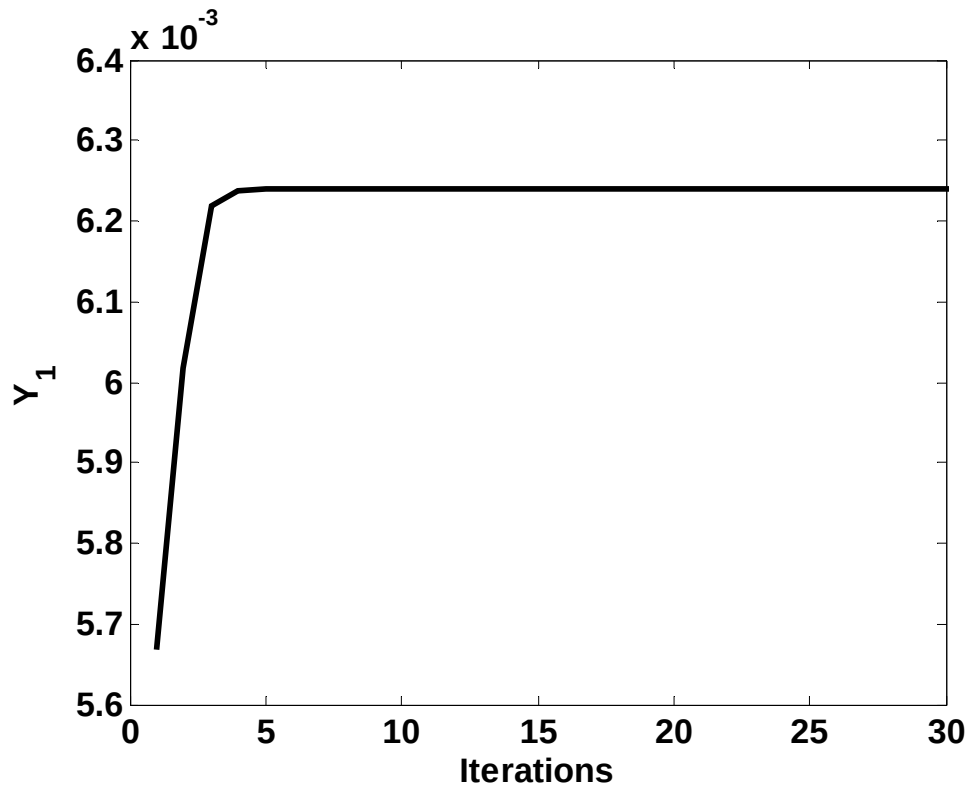

Fig. 4. Variation of enzyme activities for Case study 1 during the modified iterative IOM approach.

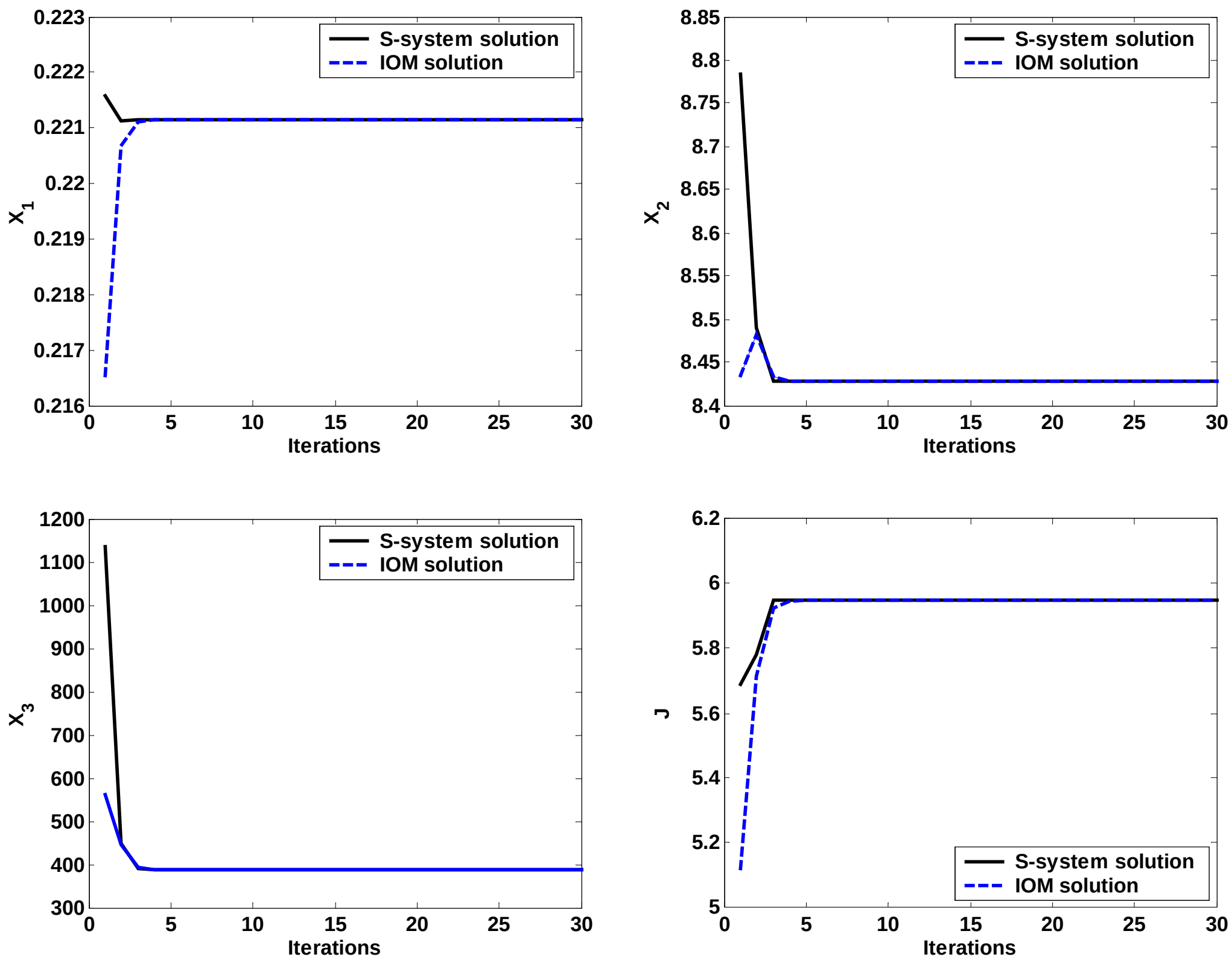

Fig. 5. Variation of metabolite concentrations and optimization index for Case study 1 during the modified iterative IOM approach.

Now we will test the quality of the above achieved new steady-state representation in terms of its stability, robustness and dynamic behavior. By solving the characteristic equation of the matrix (10), we can obtain the following eigenvalues: -0.906241, -0.026179 and -0.006559. This implies that the optimum steady-state is locally stable.

The results of the sensitivity analysis are shown in Fig. 6. Since $S(X_i, \alpha_j) = -S(X_i, \beta_j)$, Fig. 6A only shows the absolute values of sensitivities $S(X_i, \alpha_j)$. Among a total of 18 values, 15 are less than 1 and the remaining are below 3.3. Fig. 6B addresses the concentration and flux logarithmic gains. Here, $X_3$ is the most sensitive variable, and $Y_2$, $Y_8$ and $Y_9$ play an important role in the biochemical system while $Y_1$ is the determinant parameter for the tryptophan level. The influence of the kinetic orders on the metabolite concentrations and the fluxes are illustrated in Fig. 6C, where $C_P$ are given in Table 4. The variable $X_3$ exhibits the most sensitive to changes in kinetic orders while $S(X_3, h'_{3,1}) = 16.15$ and $S(X_3, h'_{3,5}) = -16.19$ are the highest sensitivities.

The dynamic response curves to a twofold increase in tryptophan concentration are plotted in Fig. 7. It can be seen that the increased tryptophan level does not have a significant effect on the mRNA concentration $X_1$ and the enzyme concentration $X_2$. The metabolites $X_1$ and $X_2$ exhibit negligible deviations from the optimum steady-state (below 0.1%). The tryptophan concentration $X_3$ shows a rapid initial decrease and asymptotically returns to within 5% of its optimum steady-state value after about 7 hours.

From the above given discussion on the analysis in local stability, robustness and dynamic behavior of the steady-state, we can conclude that the S-system structure provides a reasonably robust model description of the pathway at the optimum steady-state.

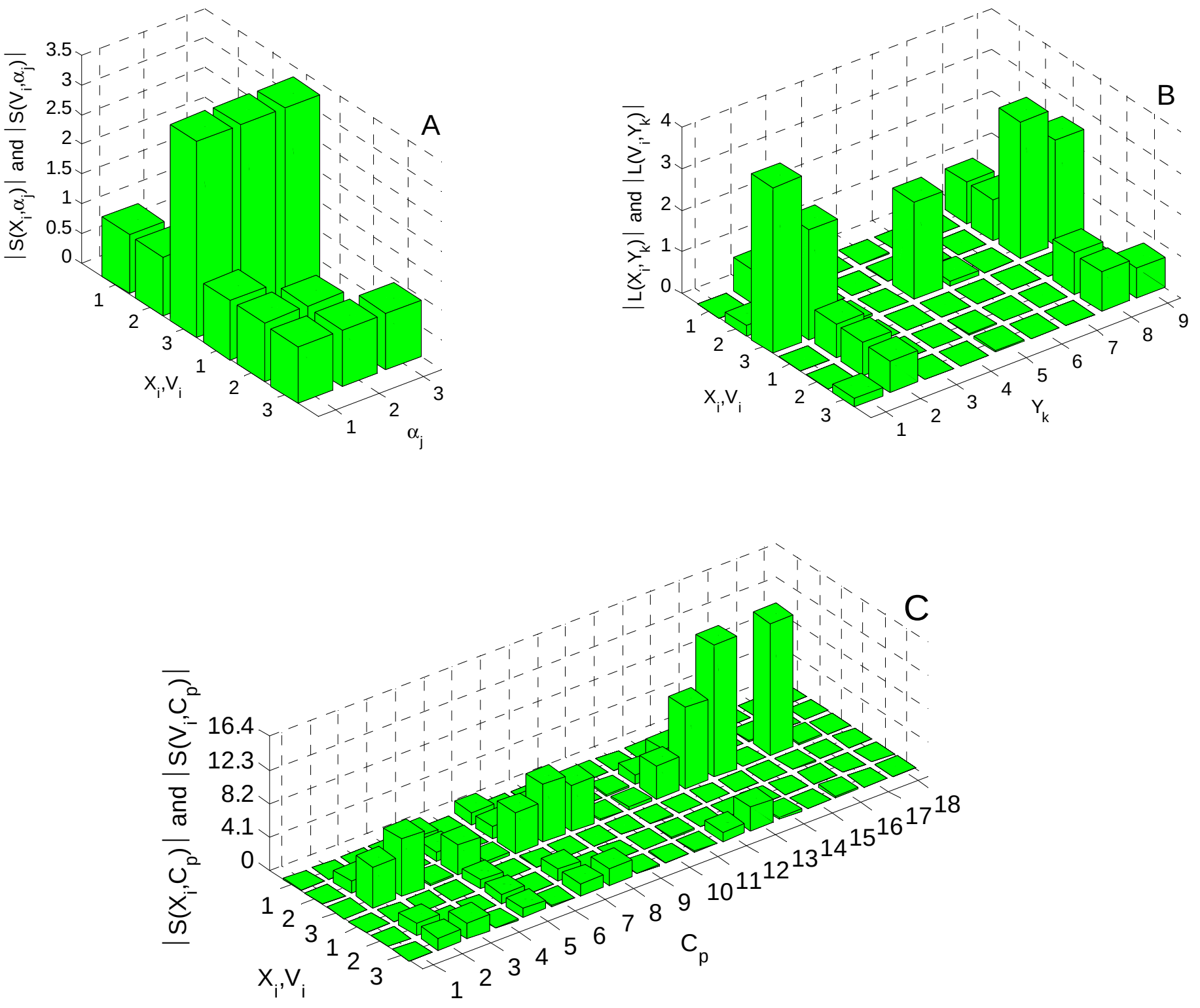

Fig. 6. Sensitivities of metabolite concentrations and fluxes with respect to changes in rate constants, kinetic orders and enzyme activities at the optimum steady-state obtained by using the modified iterative IOM approach. The 3D bar charts display the absolute magnitudes of the sensitivities and logarithmic gains. Panel A: Rate constant sensitivities. Panel B: Logarithmic gains. Panel C: Kinetic order sensitivities.

Table 4
Assignments of the kinetic orders

| $C_1$ | $C_2$ | $C_3$ | $C_4$ | $C_5$ | $C_6$ | $C_7$ | $C_8$ | $C_9$ | $C_{10}$ | $C_{11}$ | $C_{12}$ | $C_{13}$ | $C_{14}$ | $C_{15}$ | $C_{16}$ | $C_{17}$ | $C_{18}$ |
|---|---|---|---|---|---|---|---|---|---|---|---|---|---|---|---|---|---|
| $g_{1,3}$ | $g_{2,1}$ | $g_{3,2}$ | $g_{3,3}$ | $g'_{1,2}$ | $g'_{3,3}$ | $h_{1,1}$ | $h_{2,2}$ | $h_{3,3}$ | $h'_{1,1}$ | $h'_{1,8}$ | $h'_{2,1}$ | $h'_{2,9}$ | $h'_{3,1}$ | $h'_{3,4}$ | $h'_{3,5}$ | $h'_{3,6}$ | $h'_{3,7}$ |

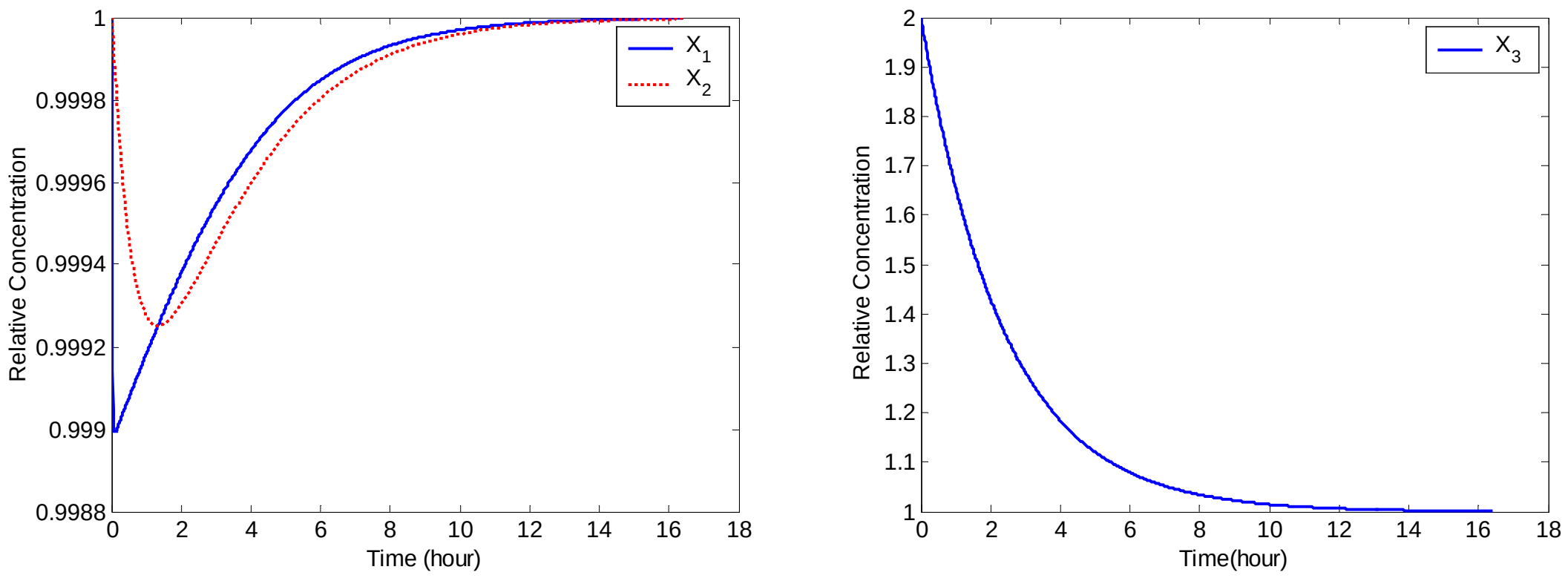

Fig. 7. Dynamic system response to a twofold increase in tryptophan concentration ($X_3$). At initial time $X_3$ is increased to twice its optimum steady-state.

## 5. Case study 2: Maximization of ethanol production in *Saccharomyces cerevisiae*

The anaerobic fermentation pathway from glucose to ethanol, glycerol and carbohydrates in the yeast *Saccharomyces cerevisiae* has been the subject of numerous studies over the years. The experimental model of this pathway in a suspended cell culture at pH 4.5 was originally proposed by Galazzo and Bailey (1990, 1991) and transformed into an S-system and studied for ethanol production maximization by Curto et al. (1995) and Torres et al. (1997). The simplified pathway is depicted graphically in Fig. 8 and its model is described as follows (see Galazzo and Bailey, 1990, 1991):

$$\frac{dX_1}{dt} = V_{\text{in}} - V_{\text{HK}} \tag{52}$$

$$\frac{dX_2}{dt} = V_{\text{HK}} - V_{\text{PFK}} - V_{\text{Pol}} \tag{53}$$

$$\frac{dX_3}{dt} = V_{\text{PFK}} - V_{\text{GAPD}} - 0.5V_{\text{Gol}} \tag{54}$$

$$\frac{dX_4}{dt} = 2V_{\text{GAPD}} - V_{\text{PK}} \tag{55}$$

$$\frac{dX_5}{dt} = 2V_{\text{GAPD}} + V_{\text{PK}} - V_{\text{HK}} - V_{\text{Pol}} - V_{\text{PFK}} - V_{\text{ATPase}} \tag{56}$$

where $X_i$ represent the following intermediate metabolite concentrations: $X_1$ is the intracellular glucose concentration, $X_2$ represents glucose 6-phosphate, $X_3$ codes for fructose 1,6-diphosphate, $X_4$ is phosphoenolpyruvate, and $X_5$ represents ATP. The indexed quantities $V$ represent the following fluxes: $V_{\text{in}}$ denotes the sugar transport into the cells, $V_{\text{HK}}$ summarizes all hexokinases, $V_{\text{PFK}}$ is the phosphofructokinase reaction, $V_{\text{GADP}}$ represents glyceraldehyde 3-phosphate dehydrogenase, $V_{\text{PK}}$ represents pyruvate kinase, $V_{\text{Pol}}$ describes glycogen synthetase, $V_{\text{Gol}}$, the glycerol 3-phosphate dehydrogenase is proportional to $V_{\text{PK}}$, and $V_{\text{ATPase}}$ summarizes collectively the use of ATP. The following flux rates in Galazzo and Bailey's model are taken within the Michaelis-Menten formalism:

$$V_{\mathrm{in}} = Y_1 - 3.7X_2 \tag{57}$$

$$V_{\mathrm{HK}} = \frac{Y_2}{\dfrac{6.2\times10^{-4}}{X_1X_5} + \dfrac{0.11}{X_1} + \dfrac{0.1}{X_5} + 1} \tag{58}$$

$$V_{\mathrm{PFK}} = \frac{50Y_3X_2X_5R_1}{R_1^2 + 3342L_1^2T_1^2} \tag{59}$$

$$R_1 = 1 + 0.3X_2 + 16.67X_5 + 50X_2X_5 \tag{60}$$

$$L_1 = \frac{1+0.76\mathrm{AMP}}{1+40\mathrm{AMP}} \tag{61}$$

$$T_1 = 1 + 1.5\times10^{-4}X_2 + 16.67X_5 + 0.0025X_2X_5 \tag{62}$$

$$\mathrm{ADP} = \frac{1}{2}(\sqrt{12X_5 - 3X_5^2} - X_5) \tag{63}$$

$$\mathrm{AMP} = 3 - X_5 - \mathrm{ADP} \tag{64}$$

$$V_{\mathrm{Pol}} = \frac{1.1Y_6}{\left(1+\left(\dfrac{2}{X_2}\right)^{8.25}\right)\left(\dfrac{1.1}{0.7X_2} + 2.43\right)} \tag{65}$$

$$V_{\mathrm{GAPD}} = \frac{Y_4}{1 + \dfrac{0.25}{X_3} + \dfrac{0.18}{\mathrm{NAD}^+}\left(1 + \dfrac{\mathrm{AMP}}{1.1} + \dfrac{\mathrm{ADP}}{1.5} + \dfrac{X_5}{2.5}\right)\left(1 + \dfrac{0.25}{X_3}\left(1 + \dfrac{\mathrm{NADH}}{0.0003}\right)\right)} \tag{66}$$

$$\mathrm{NAD}^+ = \frac{2}{Y_9 + 1} \tag{67}$$

$$\mathrm{NADH} = \frac{2Y_9}{Y_9 + 1} \tag{68}$$

$$V_{\mathrm{PK}} = \frac{Y_5X_4\mathrm{ADP}(2.519R_2 + 0.656T_2L_2^2)}{1.0832(R_2^2 + 164.084L_2^2T_2^2)} \tag{69}$$

$$R_2 = 1 + 125.94X_4 + 0.2\mathrm{ADP} + 2.519X_4\mathrm{ADP} \tag{70}$$

$$T_2 = 1 + 0.02X_4 + 0.2\mathrm{ADP} + 0.004X_4\mathrm{ADP} \tag{71}$$

$$L_2 = \frac{1+0.05X_3}{1+5X_3} \tag{72}$$

$$V_{\mathrm{Gol}} = \frac{Y_7}{Y_5}V_{\mathrm{PK}} \tag{73}$$

$$V_{\mathrm{ATPase}} = Y_8X_5 \tag{74}$$

The performance index describing the rate of ethanol production is given directly by the flux through the pyruvate kinase, $V_{\mathrm{PK}}$. The resulting optimization problem (Torres et al., 1997) is as follows:

$$\max \quad J = V_{\mathrm{PK}}$$

subject to satisfying:

$$\begin{aligned} &V_{\mathrm{in}} - V_{\mathrm{HK}} = 0 \\ &V_{\mathrm{HK}} - V_{\mathrm{PFK}} - V_{\mathrm{Pol}} = 0 \\ &V_{\mathrm{PFK}} - V_{\mathrm{GAPD}} - 0.5V_{\mathrm{Gol}} = 0 \\ &2V_{\mathrm{GAPD}} - V_{\mathrm{PK}} = 0 \end{aligned} \tag{75}$$

$$2V_{\mathrm{GAPD}} + V_{\mathrm{PK}} - V_{\mathrm{HK}} - V_{\mathrm{Pol}} - V_{\mathrm{PFK}} - V_{\mathrm{ATPase}} = 0$$
$$0.8X_{i0} \le X_i \le 1.2X_{i0} \qquad i = 1,2,3,4,5$$
$$Y_{k0} \le Y_k \le 50Y_{k0} \qquad k = 1,2,3,4,5,8$$
$$V_{\mathrm{PK}} \le 2V_{\mathrm{in}}$$
$$(Y_6, Y_7, Y_9) = (14.31, 203, 0.042)$$

This is a nonlinear optimization problem with complex constrains.

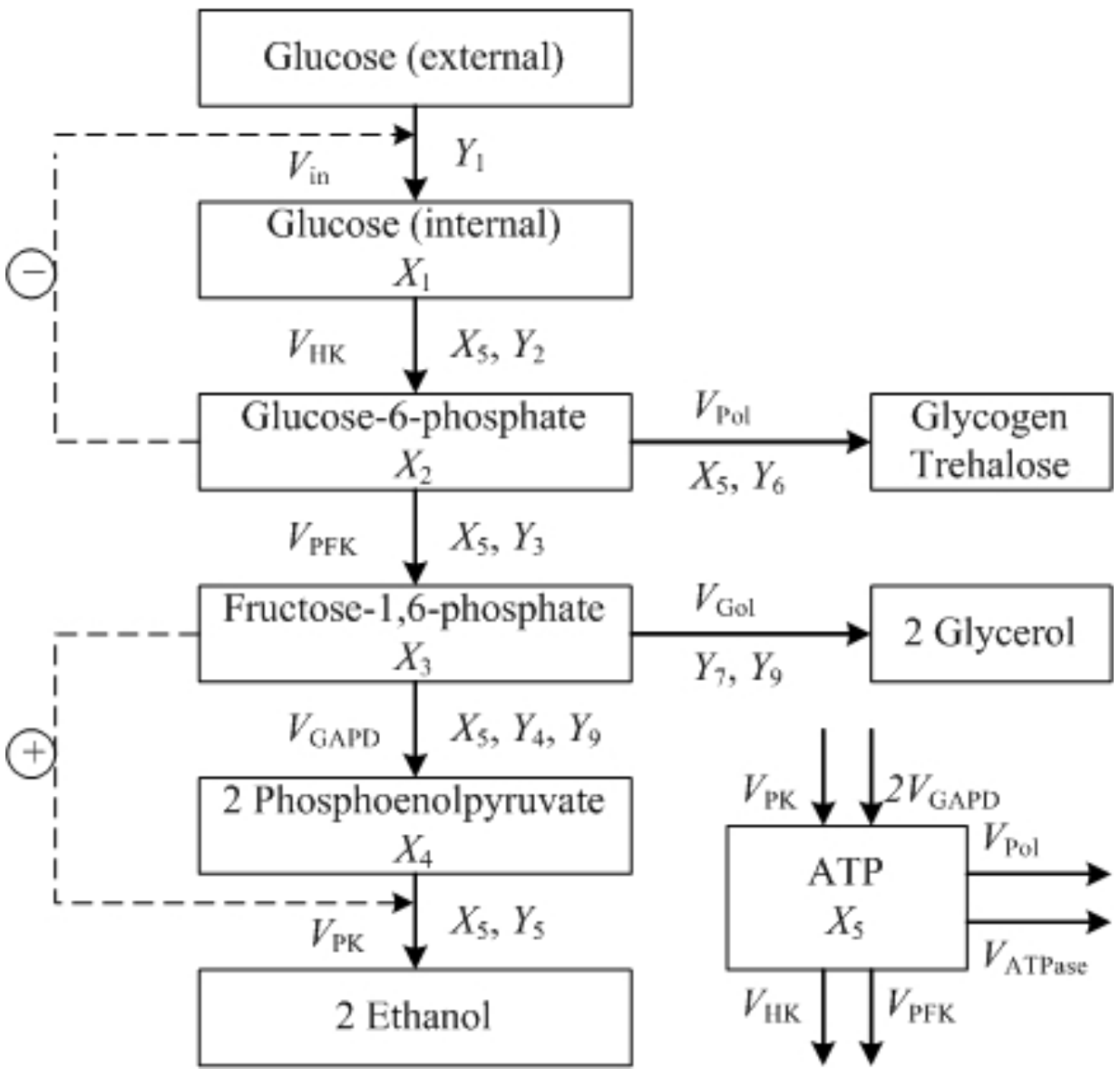

Fig. 8 Anaerobic fermentation pathway in *Saccharomyces cerevisiae*

At the basal steady-state (see Table 5), the dynamical model of the pathway is firstly transformed into an S-system by Curto et al. (1995). Here the S-system formalism is written as:

$$\frac{dX_1}{dt} = 0.8122X_2^{-0.2344}Y_1 - 2.8661X_1^{0.7464}X_5^{0.0244}Y_2 \tag{76}$$

$$\frac{dX_2}{dt} = 2.8661X_1^{0.7464}X_5^{0.0244}Y_2 - 0.5239X_2^{0.7388}X_5^{-0.3937}Y_3^{0.9991}Y_6^{0.0009} \tag{77}$$

$$\frac{dX_3}{dt} = 0.5231X_2^{0.7318}X_5^{-0.3941}Y_3 - 0.0148X_3^{0.5843}X_4^{0.0297}X_5^{0.119}Y_4^{0.9443}Y_7^{0.0557}Y_9^{-0.5749} \tag{78}$$

$$\frac{dX_4}{dt} = 0.0221X_3^{0.6159}X_5^{0.1308}Y_4Y_9^{-0.6088} - 0.0946X_3^{0.0499}X_4^{0.533}X_5^{-0.0822}Y_5 \tag{79}$$

$$\frac{dX_5}{dt} = 0.0914X_3^{0.3329}X_4^{0.2665}X_5^{0.0243}Y_4^{0.5}Y_5^{0.5}Y_9^{-0.3044} - 3.2105X_1^{0.1978}X_2^{0.1958}X_5^{0.3722}Y_2^{0.265}Y_3^{0.2648}Y_6^{0.0002}Y_8^{0.47} \tag{80}$$

After substituting the power-law terms for $V_{\mathrm{in}}$ and $V_{\mathrm{PK}}$ from Eq. (52) and (55), the nonlinear constraint $V_{\mathrm{PK}} \le 2V_{\mathrm{in}}$ becomes

$$0.0946X_3^{0.0499}X_4^{0.533}X_5^{-0.0822}Y_5 \le 1.6244X_2^{-0.2344}Y_1 \tag{81}$$

The simulation experiments of optimization problem (75) using standard and modified iterative IOM approach were performed. The following algorithm parameters were assumed in the modified method: $\theta_1 = 1$, $\theta_2 = \theta_3 = \theta_4 = 0.8$, $\eta_{1i}^0 = \eta_{2i}^0 = 0.1$, $\eta_3^0 = 0.1$. The results shown in Figs. 9 and 13 present trajectories generated by the algorithms starting from the basal steady-state given in Table 5. It can be seen that both optimization strategies yield the consistent S-system and IOM solutions respectively with a rate of ethanol production increased more than 64.8 times its basal steady-state (see Table 5 and 6). However, the modified iterative IOM approach shows a rapid convergence behavior and needs less iteration to achieve the optimum steady-state than its original version. Compared with the optimization results obtained by Torres et al. (1997), both iterative IOM methods generate a much higher rate of ethanol production than the direct IOM approach (see Table 7). These conclusions clearly show the tractability and effectiveness of the modified iterative IOM algorithm in handling large-scale biochemical systems with nonlinear constraints.

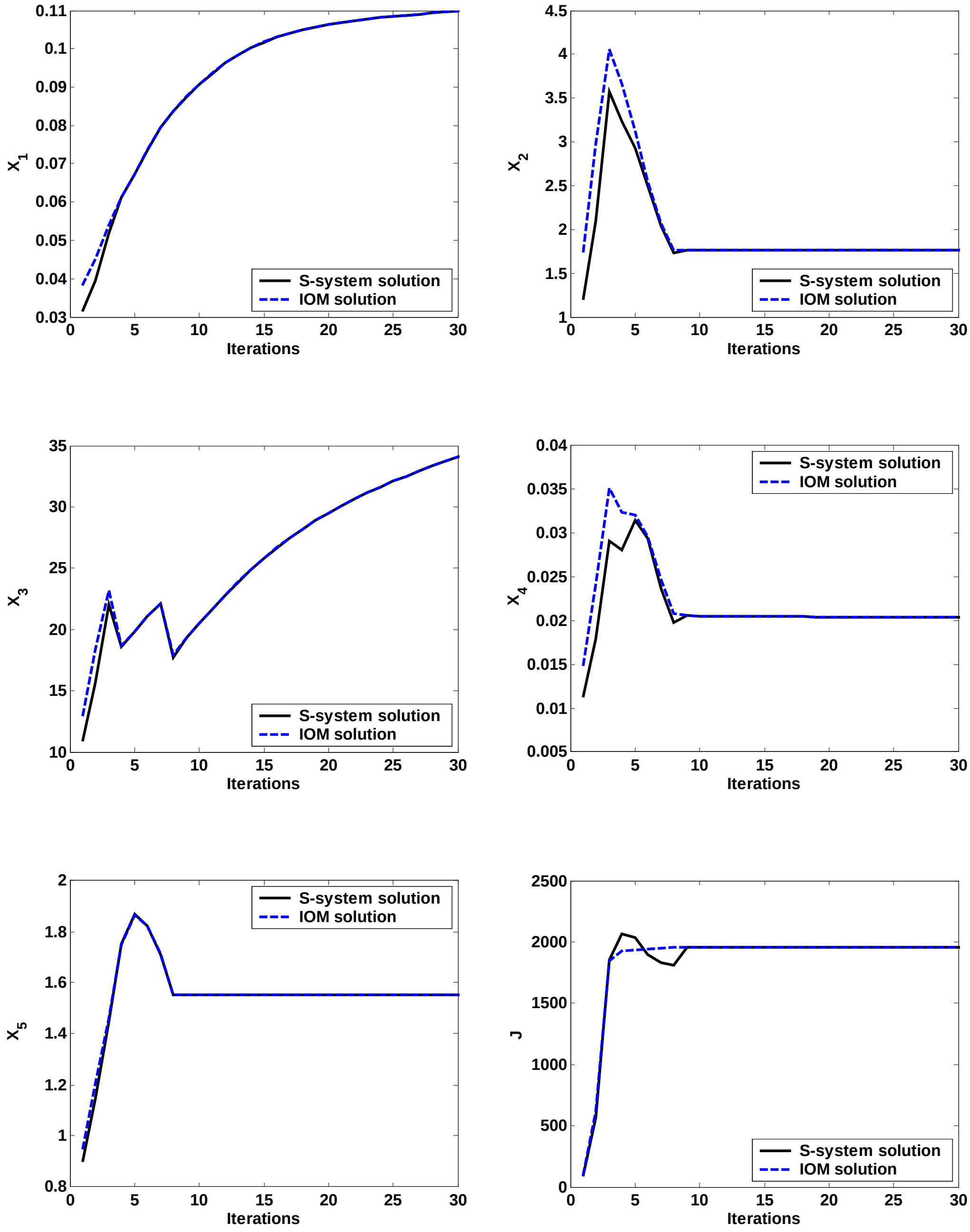

Fig. 9. Variation of metabolite concentrations and optimization index for Case study 2 during the standard iterative IOM approach.

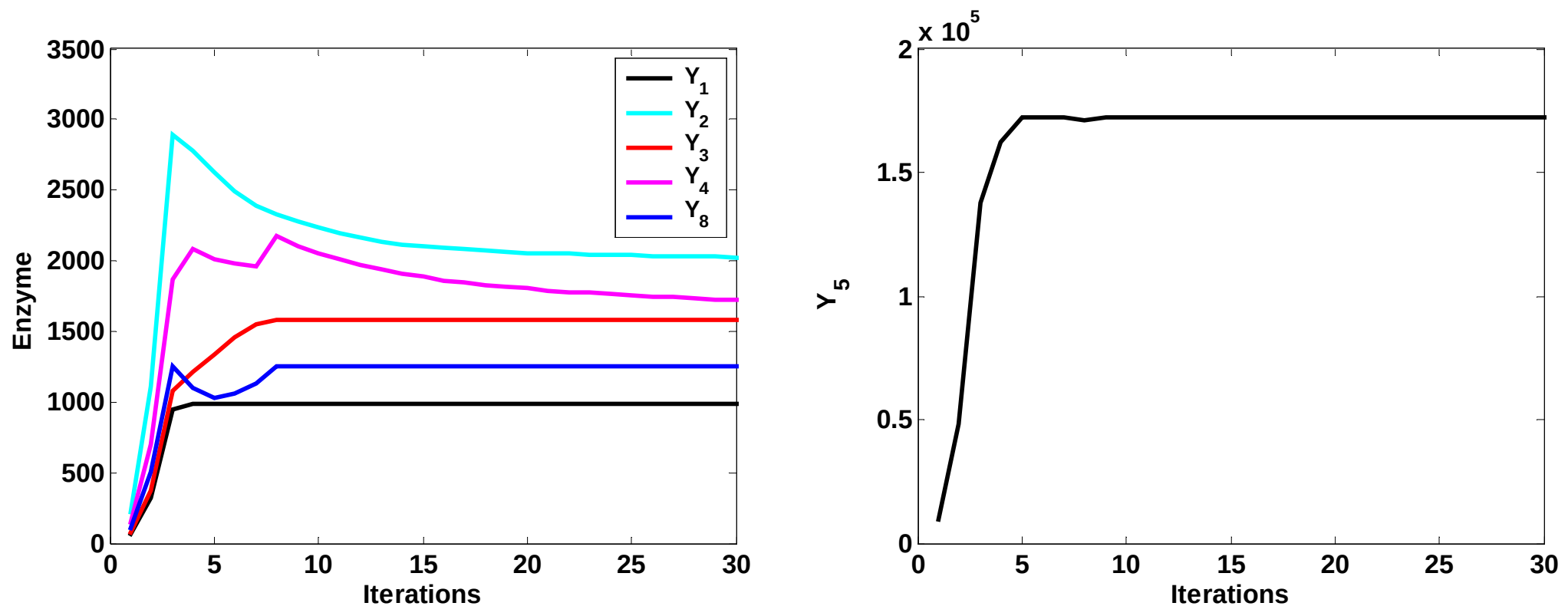

Fig. 10. Variation of enzyme activities for Case study 2 during the standard iterative IOM approach.

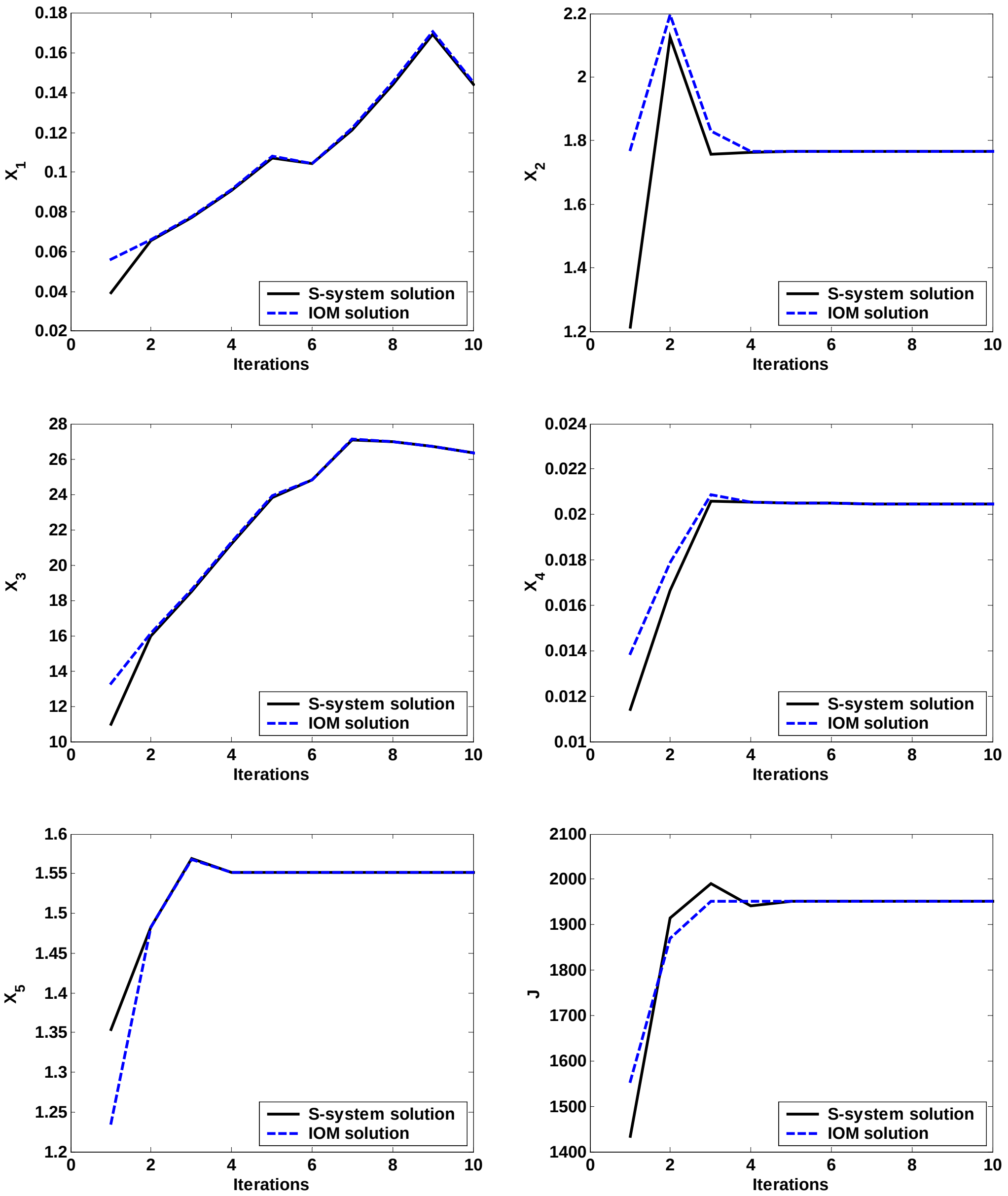

Fig. 11. Variation of metabolite concentrations and optimization index for Case study 2 during the modified iterative IOM approach.

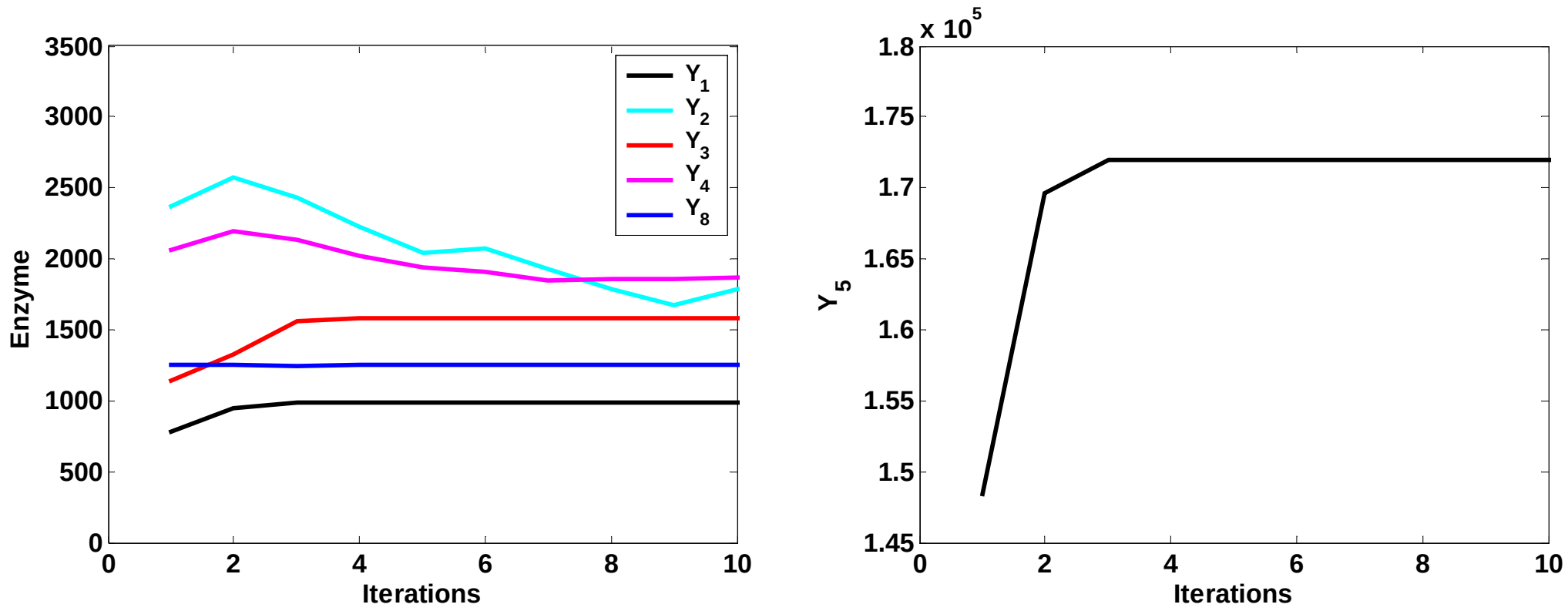

Fig. 12. Variation of enzyme activities for Case study 2 during the modified iterative IOM approach.

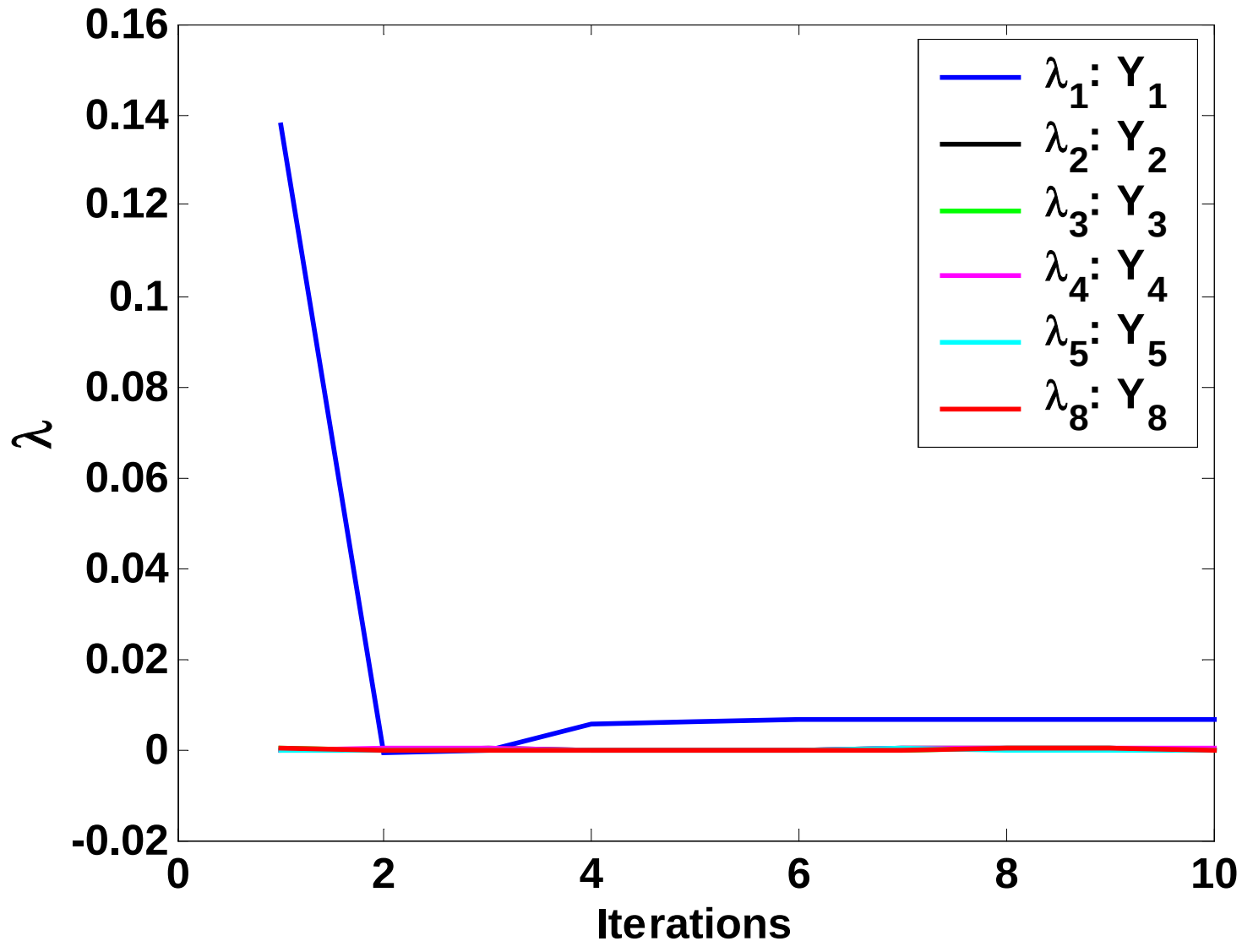

Fig. 13. Variation of Lagrangian multipliers for Case study 2 during the modified iterative IOM approach.

Table 5
Optimal solutions of Case study 2 obtained by using the standard iterative IOM approach

| Variables | Basal steady-state | Optimized solutions (15 iterations) | |
|---|---|---|---|
| | | S-system | IOM |
| $X_1$ | 0.0345 | 2.951 $(X_1)_0$ | 2.951 $(X_1)_0$ |
| $X_2$ | 1.0111 | 1.745 $(X_2)_0$ | 1.745 $(X_2)_0$ |
| $X_3$ | 9.1437 | 2.823 $(X_3)_0$ | 2.824 $(X_3)_0$ |
| $X_4$ | 0.0095 | 2.155 $(X_4)_0$ | 2.155 $(X_4)_0$ |
| $X_5$ | 1.1278 | 1.376 $(X_5)_0$ | 1.376 $(X_5)_0$ |
| $Y_1$ | 19.7 | 985 | 985 |
| $Y_2$ | 68.5 | 2102.481 | 2102.481 |
| $Y_3$ | 31.7 | 1585 | 1585 |
| $Y_4$ | 49.9 | 1883.4038 | 1883.4038 |
| $Y_5$ | 3440 | 172000 | 172000 |
| $Y_8$ | 25.1 | 1255 | 1255 |
| $J$ | 30.1124 | 64.828 $(J)_0$ | 64.829 $(J)_0$ |

Table 6
Optimal solutions Case study 2 obtained by using the modified iterative IOM approach

| Variables | Basal steady-state | Optimized solutions (6 iterations) | |
|---|---|---|---|
| | | S-system | IOM |
| $X_1$ | 0.0345 | $3.028\,(X_1)_0$ | $3.028\,(X_1)_0$ |
| $X_2$ | 1.0111 | $1.745\,(X_2)_0$ | $1.745\,(X_2)_0$ |
| $X_3$ | 9.1437 | $2.714\,(X_3)_0$ | $2.715\,(X_3)_0$ |
| $X_4$ | 0.0095 | $2.156\,(X_4)_0$ | $2.156\,(X_4)_0$ |
| $X_5$ | 1.1278 | $1.376\,(X_5)_0$ | $1.376\,(X_5)_0$ |
| $Y_1$ | 19.7 | 985 | 985 |
| $Y_2$ | 68.5 | 2075.2523 | 2075.2523 |
| $Y_3$ | 31.7 | 1585 | 1585 |
| $Y_4$ | 49.9 | 1909.9131 | 1909.9131 |
| $Y_5$ | 3440 | 172000 | 172000 |
| $Y_8$ | 25.1 | 1255 | 1255 |
| $J$ | 30.1124 | $64.829\,(J)_0$ | $64.829\,(J)_0$ |

Table 7
Optimal solutions Case study 2 obtained by using the direct IOM approach

| Variables | Basal steady-state | Optimized solutions | |
|---|---|---|---|
| | | S-system | IOM |
| $X_1$ | 0.0345 | $0.919\,(X_1)_0$ | $1.116\,(X_1)_0$ |
| $X_2$ | 1.0111 | $1.2\,(X_2)_0$ | $1.733\,(X_2)_0$ |
| $X_3$ | 9.1437 | $1.2\,(X_3)_0$ | $1.429\,(X_3)_0$ |
| $X_4$ | 0.0095 | $1.2\,(X_4)_0$ | $1.575\,(X_4)_0$ |
| $X_5$ | 1.1278 | $0.8\,(X_5)_0$ | $0.846\,(X_5)_0$ |
| $Y_1$ | 19.7 | 62.8387 | 62.8387 |
| $Y_2$ | 68.5 | 224.2612 | 224.2612 |
| $Y_3$ | 31.7 | 77.6179 | 77.6179 |
| $Y_4$ | 49.9 | 148.7649 | 148.7649 |
| $Y_5$ | 3440 | 9850.9287 | 9850.9287 |
| $Y_8$ | 25.1 | 108.5357 | 108.5357 |
| $J$ | 30.1124 | $3.159\,(J)_0$ | $3.59\,(J)_0$ |

Next, we assess the quality of S-system representation at the optimum steady-state achieved by the modified IOM method. By solving the characteristic equation of the matrix (10), we can obtain the following eigenvalues: -31077.76, -4680.565, -424.0144+394.5391*i*, -424.0144-394.5391*i* and -17.87985. This indicates local stability of the pathway model.

The sensitivities of rate constants for both intermediate metabolites and fluxes are all below 6. Among a total of 100 values, 8 range from 4 to 6 and most of them (66%) are below 1. The variable $X_4$ (phosphoenolpyruvate) appears to be the most sensitive to changes in the rate constants. The situation is rather similar with regard to the kinetic order sensitivities. Of the 115 sensitivities with respect to metabolites, most of them (73%) are below 1, with 5 bigger ones (absolute values from 4 to 7) being related to $X_4$ (phosphoenolpyruvate). In regard to the kinetic order sensitivities for the fluxes, from a total number of 115 values, most of them (90%) are below 1 and the remaining are never bigger than 1.13273. The logarithmic gains with respect to metabolites and fluxes are presented in Table 8 and 9, respectively. It can be seen that Glucose transport ( $X_6$ ) has the strongest effect on the system. The highest value in magnitude is

$L(X_4, X_6) = 4.42486$.

We have simulated two experiments to observe the dynamic system response after an increase in $X_1$ and $X_{14}$. The dynamic response curves to a tenfold increase in $X_1$ (intracellular glucose concentration) are plotted in Fig. 14 (a). It can be observed that the system quickly returns to the predisturbance steady-state with about 0.2 minutes. Fig. 14 (b) illustrates the time course of the system after a twofold increase in $X_{14}$ ([NADH]/[NAD$^+$] ratio). It clearly shows the system quickly reach a new steady-state with about 0.6 minutes.

The above analysis of stability, sensitivities, gains and dynamics demonstrates that the S-system model gives a reasonably robust description of the pathway at the optimum steady-state.

Table 8
Logarithmic gains of the S-system with respect to metabolites

| Parameters | $X_1$ | $X_2$ | $X_3$ | $X_4$ | $X_5$ |
|---|---|---|---|---|---|
| $Y_1$ | 1.91077 | 2.67278 | 2.36420 | 4.42486 | 0.94215 |
| $Y_2$ | -2.00762 | -0.00010 | 0.00004 | -0.00014 | -0.00009 |
| $Y_3$ | 0.01904 | -1.59035 | 0.05336 | 0.12853 | 0.03640 |
| $Y_4$ | 0.00003 | -0.00045 | -2.78750 | 0.03740 | -0.00041 |
| $Y_5$ | -0.00017 | 0.00238 | 0.00269 | -3.14450 | 0.00216 |
| $Y_6$ | 0.00022 | -0.00483 | -0.00261 | -0.00755 | -0.00245 |
| $Y_7$ | 0.00019 | -0.00259 | -0.00245 | -0.00718 | -0.00235 |
| $Y_8$ | 0.07754 | -1.07685 | 0.37226 | -1.43143 | -0.97541 |
| $Y_9$ | -0.00002 | 0.00022 | 0.99687 | -0.01320 | 0.00020 |

Table 9
Logarithmic gains of the S-system with respect to fluxes

| Parameters | $V_1$ | $V_2$ | $V_3$ | $V_4$ | $V_5$ |
|---|---|---|---|---|---|
| $Y_1$ | 0.98209 | 0.98209 | 0.96173 | 0.96173 | 0.96173 |
| $Y_2$ | 0 | 0 | 0 | 0 | 0 |
| $Y_3$ | 0.01066 | 0.01066 | 0.02353 | 0.02353 | 0.02353 |
| $Y_4$ | 0 | 0 | 0.00001 | -0.00020 | -0.00020 |
| $Y_5$ | -0.00002 | -0.00002 | -0.00004 | 0.00123 | 0.00107 |
| $Y_6$ | 0.00003 | 0.00003 | -0.00123 | -0.00123 | -0.00123 |
| $Y_7$ | 0.00002 | 0.00002 | 0.00004 | -0.00116 | -0.00116 |
| $Y_8$ | 0.00721 | 0.00721 | 0.01595 | 0.01603 | 0.01591 |
| $Y_9$ | 0 | 0 | 0 | 0.00010 | 0.00010 |

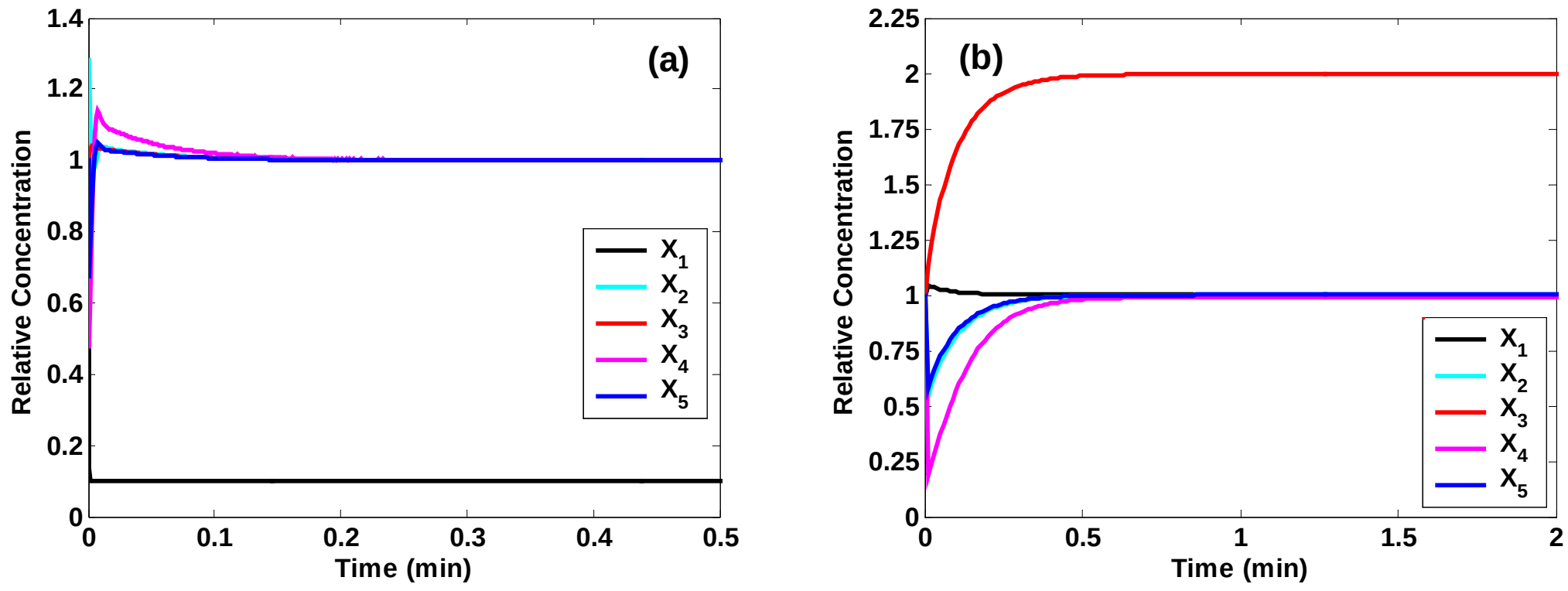

Fig. 14. Dynamic system response to perturbations. Panel (a): at time 0, the intracellular glucose concentration ($X_1$) is increased to tenfold its optimum steady-state. Panel (b): at time 0, the ratio of [NADH]/[NAD$^+$] ($X_{14}$) is increased to twofold its steady-state.

## 6. Case study 3: Optimization of a system with multiple steady-states

In this example, a biochemical system with multiple steady-states (Chang and Sahinidis, 2005) is examined. The diagram of the metabolic pathway is shown in Fig. 15 with a constant influx and the dynamics of this system are:

$$\frac{dX_1}{dt} = F + Y_1 X_3^3 - Y_2 X_1 \tag{82}$$

$$\frac{dX_2}{dt} = Y_2 X_1 - Y_3 X_2 \tag{83}$$

$$\frac{dX_3}{dt} = Y_3 X_2 - Y_4 X_3 \tag{84}$$

where $X_1$, $X_2$ and $X_3$ denote metabolite concentrations, while $Y_1$, $Y_2$, $Y_3$ and $Y_4$ denote enzyme activities.

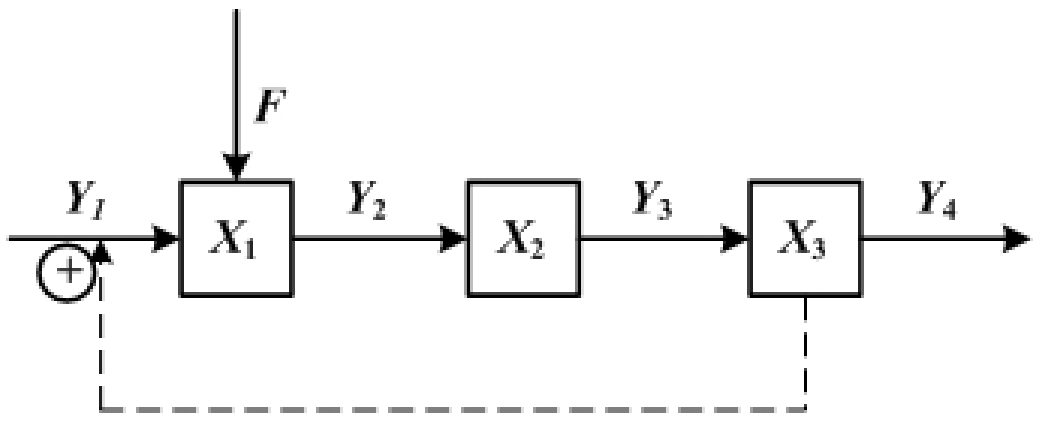

Fig. 15 Metabolite pathway of Case study 3.

At steady-state, (82)-(84) reduce to the following equations:

$$X_1 = \frac{Y_4}{Y_2} X_3 \tag{85}$$

$$X_2 = \frac{Y_4}{Y_3} X_3 \tag{86}$$

$$X_3^3 - \frac{Y_4}{Y_1} X_3 + \frac{F}{Y_1} = 0 \tag{87}$$

which determine the steady-state solutions to biochemical system (82)-(84). The distribution of roots of cubic equation (87) depends upon both the sign of the discriminant

$$D = \left(\frac{F}{2Y_1}\right)^2 - \left(\frac{Y_4}{3Y_1}\right)^3 \tag{88}$$

and relations between the roots and coefficients of this equation, which is given as: if $D > 0$, one is negative root and two are complex conjugates; if $D = 0$, one is negative root and two equal positive roots; if $D < 0$, one negative root and two different positive roots. It is clear that the case of $D > 0$ should be omitted due to no physical steady-states. Considering this and to investigate how the modified iterative IOM method can handle a system with multiple steady-states, we restrict $D$ to $D < 0$.

Let us now consider the problem of maximizing the flux $Y_1 X_3^3$:

$$\max \quad J = Y_1 X_3^3$$

subject to satisfying:

$$F + Y_1 X_3^3 - Y_2 X_1 = 0$$

$$Y_2X_1 - Y_3X_2 = 0$$
$$Y_3X_2 - Y_4X_3 = 0$$
$$0.8X_{i0} \le X_i \le 1.2X_{i0} \qquad i = 1,2,3 \tag{89}$$
$$0.2 \le Y_1 \le 5$$
$$1 \le Y_k \le 25 \qquad k = 2,3,4$$
$$\left(\frac{F}{2Y_1}\right)^2 + \tau \le \left(\frac{Y_4}{3Y_1}\right)^3$$
$$F = 4$$

where $\tau = 2/(27Y_1^2)$. The introduction of term $\tau$ guarantees that the system operates at multiple steady-states.

For this optimization problem, we choose a steady-state presented in Table 10 as the basal one. The S-system representation of the pathway at this steady-state is the following:

$$\frac{dX_1}{dt} = 5X_3^{0.6}Y_1^{0.2} - X_1Y_2 \tag{90}$$

$$\frac{dX_2}{dt} = X_1Y_2 - X_2Y_3 \tag{91}$$

$$\frac{dX_3}{dt} = X_2Y_3 - X_3Y_4 \tag{92}$$

The modified iterative IOM algorithm at the first iteration yields the solution as $(Y_1, Y_2, Y_3, Y_4) = (1.5221, 5.2058, 5.2058, 5.5116)$. Substituting these optimized parameters in the original model (82)-(84), we can get two steady-states: one is $X = (1.0712, 1.0712, 1.0118)$, which is stable; while the other is $X = (1.2528, 1.2528, 1.1833)$, which is unstable. Thus, the former steady-state can be chosen as the basal one at the second iteration. Repetition of above-mentioned process would produce a sequence of steady-state solutions, which move towards the true optimum. The corresponding trajectories in metabolite concentrations and optimization index during the modified iterative IOM approach are shown in Fig. 16. It can be seen that the proposed algorithm obtains the consistent S-system and IOM solutions with objective value 1.576. The detailed results within 2 iterations are given in Table 10. The relaxation coefficients $\theta_1$, $\theta_2$ and $\theta_3$ are selected as 1.0, 0.9 and 0.9 respectively.

Table 10
Optimal solutions of Case study 3 obtained by using the modified iterative IOM approach

| Variables | Basal steady-state | Optimized solutions (2 iterations) | |
|---|---|---|---|
| | | S-system | IOM |
| $X_1$ | 1 | $1.026\,(X_1)_0$ | $1.026\,(X_1)_0$ |
| $X_2$ | 1 | $1.026\,(X_2)_0$ | $1.026\,(X_2)_0$ |
| $X_3$ | 1 | $1.006\,(X_3)_0$ | $1.006\,(X_3)_0$ |
| $Y_1$ | 1 | 1.5472 | 1.5472 |
| $Y_2$ | 5 | 5.4361 | 5.4361 |
| $Y_3$ | 5 | 5.4361 | 5.4361 |
| $Y_4$ | 5 | 5.5417 | 5.5417 |
| $J$ | 1 | $1.576\,(J)_0$ | $1.576\,(J)_0$ |

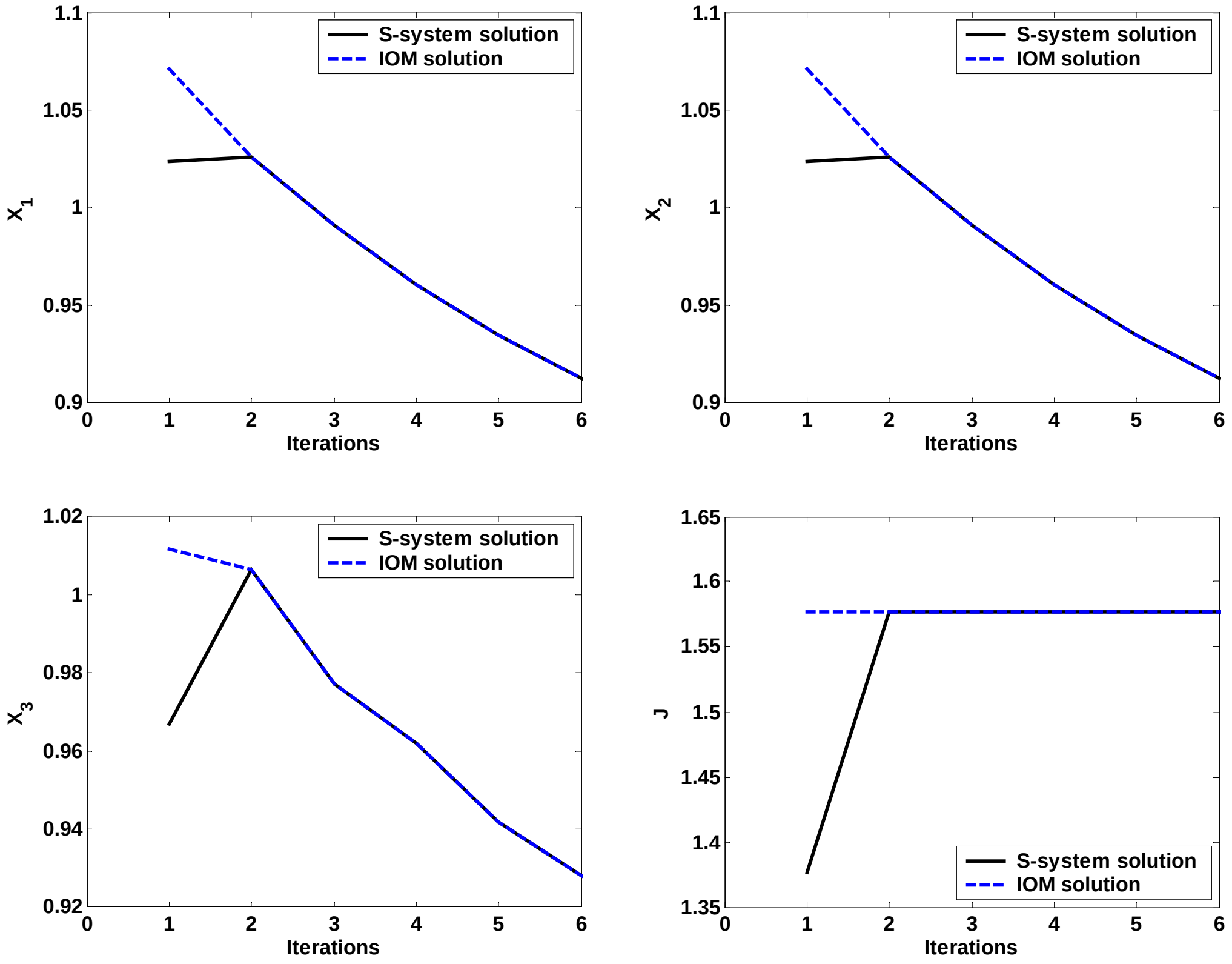

Fig. 16. Trajectories in metabolite concentrations and optimization of Case study 3 index during the modified iterative IOM approach.

## 7. Conclusions

In this paper, an algorithm for optimization of biochemical systems has been presented. Using a modification of the existing iterative IOM approach to account for differences of metabolite concentration derivatives with respect to enzyme activities between the S-system and the original model enables the modified method to achieve the correct optimal steady-state solution and ensures the modified algorithm to be implemented within the linear optimization techniques. The proposed framework has been applied to three biochemical systems. The simulation results show that the modified iterative IOM strategy rapidly and successfully maximizes the performance index required for these systems, whereas the standard iterative IOM approach either fails to obtain the correct optimum steady-state point or shows a slower evolution rate than its new version.

## Acknowledgements

We would like to acknowledge the comments and suggestions made by the anonymous referees. This work was supported by a research grant from State Science and Technology Project (No. 2001BA204B01) of P. R. China.